\documentclass[journal]{IEEEtran}
\usepackage{cite}
\usepackage{amsmath,amssymb,amsfonts}
\usepackage[pdftex]{graphicx}
\usepackage{gensymb}
\usepackage{stfloats}
\usepackage{float}
\usepackage{cite}
\usepackage[caption=false,font=footnotesize]{subfig}

\usepackage{float}
\usepackage[colorlinks=true,linkcolor=blue,citecolor=blue,urlcolor=blue]{hyperref}
\usepackage{xcolor}
\usepackage{siunitx}
\usepackage{multirow}
\usepackage{threeparttable} 

\definecolor{authorred}{rgb}{0.0, 0.0, 0.0}
\definecolor{authorred2}{rgb}{0.0, 0.0, 0.0}

\ifCLASSINFOpdf
\else
\fi

\begin{document}

\title{
  Negative Resistance Caused by Intra-Loop Coupling in Virtual-Admittance-Based Grid-Forming Control
}

\author{Jaekeun~Lee,~\IEEEmembership{Graduate~Student~Member,~IEEE,}
        Minwoo~Jeong,~\IEEEmembership{Graduate~Student~Member,~IEEE,}
        Jae-Jung~Jung,~\IEEEmembership{Senior~Member,~IEEE,}
        Xiongfei~Wang,~\IEEEmembership{Fellow,~IEEE,}
        Shenghui~Cui,~\IEEEmembership{Senior~Member,~IEEE}
} 

\markboth{ }%
{}

\maketitle

\begin{abstract}
  This paper addresses the harmonic instability problem of
  the virtual-admittance (VA)-based grid-forming control.
  It is revealed that the intra-loop coupling among
  the VA control, the inner-loop current control,
  and the voltage feedforward control
  results in an \(s^2\)-term in the equivalent output impedance of the inverter,
  which induces a negative-resistance property in the harmonic range.
  It is worth highlighting that this negative resistance is independent of the control delay.
  Consequently, this harmonic instability mechanism is fundamentally different
  from the extensively investigated cases in the literature,
  which are induced by the digital control delay of inverters.
  Then, a simple passivity-oriented damping control is proposed to mitigate
  the negative resistance arising from the intra-loop coupling.
  The method fully retains the well-established current controller and voltage feedforward,
  and does not require grid impedance information.
  Finally, experimental tests verify the theoretical findings and the effectiveness of the damping method.
\end{abstract}


\IEEEpeerreviewmaketitle

\section*{Nomenclature}
\addcontentsline{toc}{section}{Nomenclature}
\begin{IEEEdescription}[\IEEEsetlabelwidth{$Z_\text{eq,Pade}^\text{prop}(s)$}]
\item[CC]    Current controller.
\item[GFL]   Grid-following.
\item[GFM]   Grid-forming.
\item[IBR]   Inverter-based resource.
\item[IVS]   Internal voltage source.
\item[PCC]   Point of common coupling.
\item[PR]    Proportional-resonant.
\item[PVR]   Parallel virtual resistance.
\item[PWM]   Pulse-width modulation.
\item[SCR]   Short-circuit ratio.
\item[SISO]  Single-input single-output.
\item[VA]    Virtual admittance.
\item[VA-CC] Cascaded inner-loop structure of VA, CC, and VFF.
\item[VFF]   Voltage feedforward.

\item[$\mathbf{e}$]            Internal voltage source from the outer power control loop.
\item[$\mathbf{i}$]            Inverter output current.
\item[$\mathbf{i}_\text{ref}$] Current reference generated by VA.
\item[$\mathbf{v}$]            Voltage at the PCC.
\item[$\mathbf{v}_\text{inv}$] Inverter modulated output voltage.

\item[$C_\text{g}$]       Grid-side shunt capacitance.
\item[$L_\text{f}$]       Filter inductance.
\item[$L_\text{v}$]       Virtual inductance.
\item[$R_\text{c}$]       Damping resistance of $C_\text{g}$.
\item[$R_\text{v}$]       Virtual resistance (series).
\item[$R_\text{v,p}$]     Parallel virtual resistance (proposed).
\item[$n_\text{XR}$]      X/R ratio of conventional VA.

\item[$K_\text{p}$]  Proportional gain of CC.
\item[$K_\text{r}$]  Resonant gain of CC.

\item[$f_1$, $\omega_1$]              Fundamental grid frequency (Hz, rad/s).
\item[$f_\text{cc}$, $\omega_\text{cc}$] CC bandwidth (Hz, rad/s).
\item[$f_\text{cr}$]                  Critical frequency above which\\ $\Re[Z_\text{eq,0}(j\omega)]<0$.
\item[$f_D$, $\omega_D$]              Resonant frequency of $Z_\text{eq}(s)$ (Hz, rad/s).
\item[$f_\text{LC,g}$]                Grid-side LC resonant frequency.
\item[$f_\text{s}$]                   Sampling frequency.
\item[$T_\text{d}$]                   Control delay.

\item[$G_\text{cc}(s)$]              Current controller.
\item[$G_\text{d}(s)$]               Control-delay model, $e^{-sT_\text{d}}$.
\item[$Y_\text{v}(s)$]               Virtual admittance (conventional).
\item[$Z_\text{eq}(s)$]              Equivalent output impedance of VA-CC.
\item[$Z_\text{eq,0}(s)$]            Delay-absent model of $Z_\text{eq}(s)$.
\item[$Z_\text{eq,Pade}(s)$]         First-order Padé-approximated $Z_\text{eq}(s)$.
\item[$Z_\text{eq}^\text{prop}(s)$]  $Z_\text{eq}(s)$ under proposed VA design.
\item[$Z_\text{eq,0}^\text{prop}(s)$] Delay-absent model of $Z_\text{eq}^\text{prop}(s)$.
\item[$Z_\text{g}(s)$]               Grid-side impedance seen from the PCC.
\end{IEEEdescription}

\section{Introduction}

\IEEEPARstart{P}{ower} electronic devices are becoming
increasingly prevalent in modern electric power grids.
To accommodate the growing penetration of inverter-based resources (IBRs),
grid-forming (GFM) inverters are gaining significant attention
\cite{Rathnayake2021Grid, Tozak2024Modeling, Zhang2021Grid, kroposki2026unifi, NESO2023GFM,Song2022Review,musca2022grid}.
Unlike conventional grid-following (GFL) inverters,
GFM inverters exhibit voltage-source behavior
analogous to that of synchronous generators.
However, unlike synchronous generators, inverters employ fast-acting control loops.
These control loops introduce additional harmonic stability concerns,
even when the outer loop is configured to emulate voltage-source behavior.
Since the outer control loop of a GFM inverter typically exhibits slow dynamics
\cite{kroposki2026unifi, NESO2023GFM},
harmonic stability is predominantly governed by the inner control loops
\cite{Wu2023Passivity, Ravanji2023Impact, He2025Enhancing, Akhavan2023Passivity}.

Among the various types of inner loops, virtual admittance (VA) control
\cite{Rodriguez2013Control, Leon2023Grid, Tarraso2017Grid}
is widely favored
owing to its emulation of virtual inductance without a differentiator,
and its compatibility with well-established
inner current regulation strategies: the current controller (CC) 
in conjunction with the voltage feedforward (VFF).
While some works omit the VFF
\cite{imgart2026frequency, Cui2024Virtual, Huang2021Impact}
in the control loop and the stability analysis,
the VFF is a key control element in practice for rejecting arbitrary disturbances from the grid voltage
\cite{Xiong2019Physical,Li2018Capacitor,Yan2016Improved,Li2011full,Wang2021Robust}.
Therefore, this work regards the VFF as one of the essential components of VA-based GFM control.
This inner control loop structure, consisting of VA, CC, and VFF---hereafter referred to as VA-CC---enables
a GFM inverter to operate reliably under various grid conditions.
However, many works have reported harmonic instability issues associated with the VA-CC-based GFM control
\cite{Ravanji2023Impact, Gao2026Small, Li2026Analysis, Obi2025Revaluation, Zhang2025Impedance, cvetanovic2026precise, Feng2024Small, Kamalinejad2025Comprehensive, Beza2021Impact, Miranbeigi2022Passivity}.
To examine the harmonic stability, many of these analyses
typically incorporate control delay from the outset, as it has long been
regarded as the dominant cause of inverter non-passivity in the harmonic range
\cite{Harnefors2014Passivity,Harnefors2016Passivity, Zou2018Analysis,Holmes2009Design, Agbemuko2021Passivity, Harnefors2015Passivity, Harnefors2017Nyquist}.

However, beyond the well-known control-delay-induced non-passivity, 
one may further ask whether the controller structure of an inverter
can trigger harmonic instability on its own---that is,
in the absence of control delay and without invoking any nonlinear mechanism.
If the grid impedance is composed of passive elements, this question reduces to
whether the interaction among individually passive linear control loops
can render the output impedance of an inverter to be non-passive,
regardless of control delay or nonlinearity.
A linear control structure that introduces non-passivity on its own---irrespective of the control delay---should
be regarded as carrying a fundamental design shortcoming.


This paper reveals that VA-CC introduces the aforementioned type of non-passivity,
and traces the root cause.
It is shown that the intra-loop coupling among the VA, CC, and VFF
readily induces a negative-resistance property in the harmonic range,
in a control-delay-independent manner.
Specifically, this coupling gives rise to
an \(s^2\)-term in the equivalent output impedance of an inverter with VA-CC.
This \(s^2\)-term
renders the overall inner control structure non-passive,
even when the control delay can be assumed negligible.
Consequently, a VA-CC-based GFM inverter can readily trigger
harmonic instability when connected
to a grid with a capacitive shunt component.
The identified negative-resistance and harmonic-instability mechanism is explained
through a simple single-input single-output (SISO) impedance-based analysis.


Building on the identified non-passivity mechanism, i.e., the intra-loop coupling in VA-CC,
a simple passivity-oriented design is then presented as an example to address such issue.
To avoid the non-passivity arising from the intra-loop coupling,
the proposed VA is designed to be resistive in the high-frequency range.
To achieve this, a parallel virtual resistance (PVR) is added to the VA.
It is shown that simply adding a PVR with the VA allows the
overall inner control structure---comprising the proposed VA, CC, and VFF---to
ensure a positive-real property up to a sufficiently high frequency.
The simple design rule for PVR is also provided.

The authors acknowledge that many prior works
have examined the harmonic instability of VA-CC.
These studies use either
multi-input multi-output (MIMO) small-signal analysis~\cite{Gao2026Small,Zhang2025Impedance,cvetanovic2026precise,Feng2024Small,Kamalinejad2025Comprehensive, Beza2021Impact, Miranbeigi2022Passivity},
or inspection of the open-loop transfer function~\cite{Ravanji2023Impact,Obi2025Revaluation,Li2026Analysis}.
Typically, these approaches simultaneously incorporate the entire control loop,
control delay, and grid impedance.
As a result, they have not specifically identified the {control-delay-independent}
negative-resistance property from VA-CC itself.
In contrast to previous works,
this work elucidates, through a simple SISO impedance-based analysis,
how the intra-loop coupling in VA-CC
induces a negative-resistance property
in a control-delay-independent manner.{
More importantly,
this work adds a new perspective
to prior passivity-based analyses~\cite{Harnefors2014Passivity,Harnefors2016Passivity, Beza2021Impact, Miranbeigi2022Passivity, Agbemuko2021Passivity, Harnefors2015Passivity}
which regard the control delay as the dominant cause of non-passivity in the harmonic range.
In contrast, this work reveals a mechanism in which individually passive linear control loops
collectively induce a negative-resistance property,
independent of the control delay.}
Regarding the mitigation,
unlike previous works that modify the CC or VFF for stability
\cite{Ravanji2023Impact, Gao2026Small,Obi2025Revaluation}
and thereby inevitably sacrifice bandwidth and dynamic performance
of current regulation,
a VA-level passivity-enhancement method is adopted.
Therefore, the proposed method can exploit the high sampling frequency
to achieve a high CC bandwidth within the well-established
current-regulating structure.
The proposed method also differs from prior approaches based on the
open-loop transfer function \cite{Li2026Analysis, Obi2025Revaluation},
which require knowledge of grid impedance.
Owing to its passivity-oriented design,
the proposed method ensures the positive-real property up to a
sufficiently high frequency.
Thereby, it maintains stability under practical passive grid conditions,
without requiring grid impedance information.
Note that despite the passivity-enhancement method presented in this paper
provides an intuitive physical insight for design,
other methods of loop shaping can also be exploited upon a careful consideration of
the negative-resistance mechanism discovered in this work.

The remainder of this paper is organized as follows.
Section~\ref{Section:II} analyzes the identified negative resistance of VA-CC in a simplified manner.
First, the control-delay-independent harmonic instability of VA-CC is demonstrated.
The cause of the harmonic instability
is traced to the negative-resistance property arising from the \(s^2\)-term,
a consequence of the intra-loop coupling within VA-CC.
Section~\ref{Section:III} examines the detailed frequency range in which
VA-CC exhibits a negative-resistance property with the control delay taken into account.
Furthermore, it is shown that if the grid-side LC resonant frequency
lies within the examined negative-resistance region of VA-CC,
the inverter is likely to induce harmonic instability.
Section~\ref{Section:IV} introduces the mitigation method,
achieved by adding a certain amount of PVR to the VA.
Section~\ref{Section:V} verifies the analysis
and the mitigation method through experimental results.
Finally, Section~\ref{Section:VI} concludes the paper.

\section{Identification of Negative-Resistance Property from Intra-Loop Coupling within VA-CC}
\label{Section:II}


This paper focuses on the harmonic instability arising from the inner control structure---VA-CC---of
a GFM inverter, which provides a necessary condition for stability.
The underlying assumption is that the outer loops ensure slow
internal voltage source (IVS) dynamics, which is a necessary requirement for GFM operation~\cite{kroposki2026unifi, NESO2023GFM}.
Therefore, the harmonic analysis can be focused on VA-CC.

\subsection{System Description}

\begin{figure}[!t]
    \centering
    \subfloat[]{
       {\includegraphics[width=0.625\linewidth]{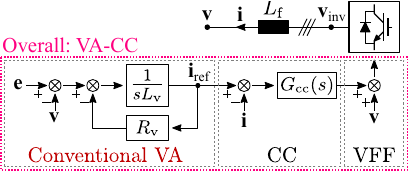}
       }
    }
    \vspace{0pt}
    \subfloat[]{\includegraphics[width=0.98\linewidth]{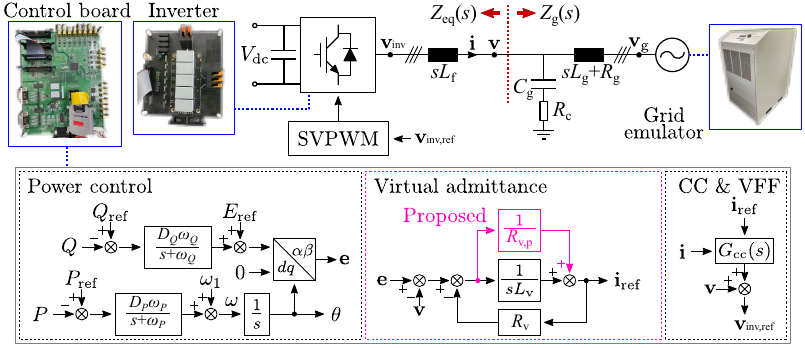}}
    \caption{(a) Control diagram VA-CC, composed of VA, CC, and VFF.
    (b) Full system under investigation. The proposed block $ 1/ R_\text{v,p} $ is omitted
    in the investigation of conventional VA-CC.
    }
    \label{fig:Diagram_VACC}
\end{figure}

The overall system is depicted in Fig.~\ref{fig:Diagram_VACC}.
Fig.~\ref{fig:Diagram_VACC}(a) depicts the conventional VA-CC, comprising VA, CC, and VFF.
Fig.~\ref{fig:Diagram_VACC}(b) depicts the overall control diagram.
An L-filter inverter is considered in this study, and the grid includes shunt capacitance at the point of common coupling (PCC).
The outer loop consists of droop control for both active and reactive power control.
Both employ a low-pass filter to ensure slow IVS dynamics.
In this paper, the droop gains and the cut-off frequencies are fixed at 0.05~p.u.
The VA loop is denoted as \(Y_\text{v}(s)\). Conventionally, it consists of a virtual resistance
and a virtual inductance, \(R_\text{v}\) and \(L_\text{v}\), respectively,
i.e., \(Y_\text{v}(s) = 1/(s L_\text{v} + R_\text{v}) \).
\(G_\text{cc}(s)\) denotes the CC.
In this paper, a proportional-resonant (PR) controller in the stationary frame is utilized~\cite{Zmood1999Stationary, Baeckeland2022Stationary}.
The transfer function is expressed as follows, where \(\omega_1\) is the fundamental angular frequency of the grid:
\begin{equation}
  \begin{aligned}
    G_\text{cc}(s) & = K_\text{p} + \frac{2 K_\text{r} \omega_1 s}{s^2 + 2 \zeta \omega_1 s + \omega_1^2}
    \\
    & \approx K_\text{p} \,\ \text{if } \omega \gg \omega_1.    
  \end{aligned}
\end{equation}
The resonant controller should not interfere with other control loops for CC-level stability,
and therefore an excessive \(K_\text{r}\) should be avoided~\cite{Harnefors2014Passivity, Harnefors2016Passivity, Holmes2009Design}.
Throughout this paper, \(K_\text{r}\) is fixed at \(0.1 K_\text{p}\), and
\(\zeta=0.001\).
The proportional gain is designed based on the CC bandwidth as \(K_\text{p} = \omega_\text{cc} L_\text{f}\),
where \(\omega_\text{cc}\) and \(f_\text{cc}\) denote the CC bandwidth in rad/s and Hz, respectively.


\begin{table}[t]
\centering
\caption{Section II Simulation Parameters: Default Values}
\label{tab:SectionII}
\begin{tabular}{c c c}
\hline
\text{Parameter} & \text{Symbol} & \text{Value} \\
\hline
Fundamental grid frequency & \(f_1\) & 60~Hz \\
Control delay & \(T_{\text{d}}\) & 0~\(\mu\)s, ideal \\
Rated apparent power & \(S_{\text{n}}\) & 3.0 kVAR \\
Grid voltage, line-to-line RMS & \(V_{\text{g,rms}}\) & 220 V \\
Grid short circuit ratio (SCR) & SCR & 4 \\
Grid X/R ratio & \(n_\text{XR,g}\) & 5 \\
Shunt capacitance & {\(C_{\text{g}}\)} & {6 \(\mu\)F} (0.036~p.u.) \\
Damping resistance & {\(R_{\text{c}}\)} & {10 m\(\Omega\)} (6.2e-4~p.u.) \\
Filter inductance & {\(L_{\text{f}}\)} & {3.4 mH} (0.079~p.u.) \\
Cut-off frequency of \(G_P(s)\), \(G_Q(s)\) & {\(\omega_{{P}}, \omega_{{Q}}\)} & \(6\pi\)~rad/s (0.05~p.u.) \\
Droop gains of the outer loop & {\(D_{{P}}, D_{{Q}}\)} & 0.05~p.u. \\
Virtual inductance & {\(L_{\text{v}}\)} & {21.4 mH} (0.5~p.u.) \\
X/R ratio of VA & {\(n_{\text{XR}}\)} & 5 \\
Control bandwidth of CC & {\(f_{\text{cc}}\)} & 1~kHz (16.667~p.u.) \\
\hline
\end{tabular}
\end{table}

\subsection{Demonstration of Harmonic Instability of VA-CC in the Absence of Control Delay}

\begin{figure}[!t]
    \centering
        \includegraphics[width=0.75\linewidth]{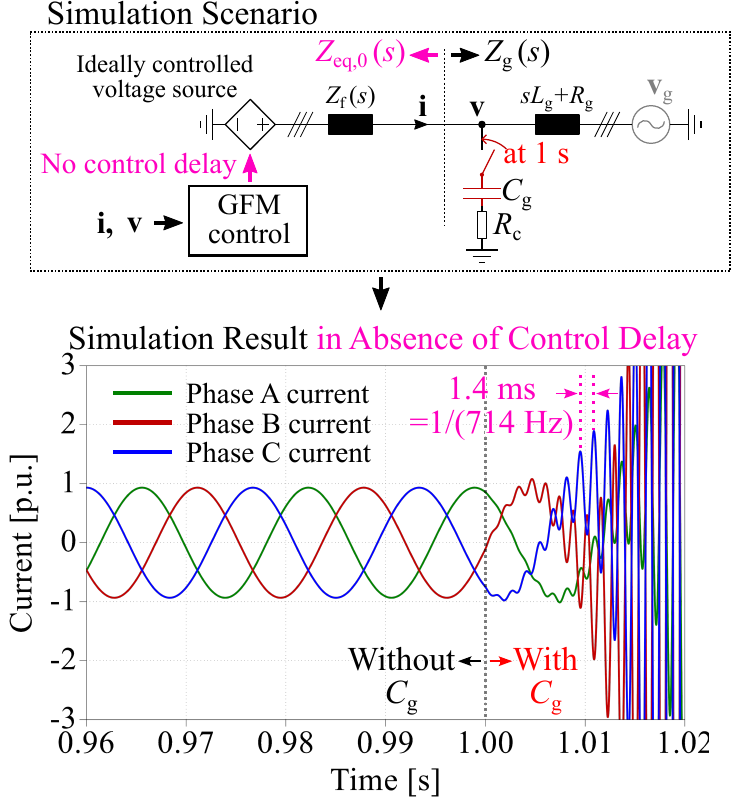}
    \caption{Simulation result of a GFM inverter modeled as an ideal controlled voltage source
    in the absence of control delay, exhibiting instability with connection of a shunt capacitor after \(t=1\)~s.}
    \label{fig:Validation2_NoDelaySimulation}
\end{figure}

To demonstrate the fundamental shortcoming of VA-CC,
an unstable simulation result in the absence of control delay is presented.
The GFM inverter is modeled as an ideally controlled,
delay-free switching-averaged voltage source
in a PLECS simulation,
as shown in Fig.~\ref{fig:Validation2_NoDelaySimulation}.
The parameters are detailed in Table~\ref{tab:SectionII}.
In the simulation, the shunt capacitor is initially disconnected,
and therefore the grid is purely inductive. Under the purely inductive grid condition,
the GFM inverter operates stably.
For a seamless voltage transition,
the capacitors are precharged and then connected to the grid at \(t=1\)~s.
Soon afterward, the output current oscillates, even with the delay-free, ideally controlled voltage source model.
The main oscillation frequency is 714~Hz.

Since the grid-side impedance consists only of passive components and
an infinite voltage source, the harmonic instability in Fig.~\ref{fig:Validation2_NoDelaySimulation} indicates that
even the \emph{delay-free} inverter is non-passive in the harmonic range.
In other words, the equivalent output impedance of the inverter exhibits a negative-resistance property
in the harmonic range,
despite the absence of control delay.

\subsection{Identification of Negative-Resistance Property from VA-CC via Simplified Output Impedance Model}

To identify the negative-resistance mechanism---that arises even without control delay---a
simplified derivation is followed.
Here, the control delay \(G_\text{d}(s)\) is considered negligible, i.e., \(G_\text{d}(s) = 1\).
In this case, the modulated output voltage, \(\mathbf{v}_\text{inv}\), is determined as:
\begin{equation}
  \mathbf{v}_\text{inv} = G_\text{cc} (s) Y_\text{v}(s) \left( \mathbf{e} - \mathbf{v} \right) - G_\text{cc} (s) \mathbf{i} + \mathbf{v},
\label{eq:Simple_Vinv}
\end{equation}
where \(\mathbf{e}\), \(\mathbf{v}\), and \(\mathbf{i}\) denote the IVS, the PCC voltage, and the inverter-side current,
respectively, all expressed as space vectors in the stationary reference frame.
Meanwhile, the relationship between the PCC voltage \(\mathbf{v}\) and \(\mathbf{v}_\text{inv}\) is determined by
the filter inductance:
\begin{equation}
  \mathbf{v} = \mathbf{v}_\text{inv} - s L_\text{f} \mathbf{i}.
\label{eq:Simple_V}
\end{equation}
In other words, the VFF holds the information of \(sL_\text{f}\).
Substituting \eqref{eq:Simple_Vinv} into \eqref{eq:Simple_V} yields
\begin{equation}
  \begin{aligned}
    \left(sL_\text{f} + G_\text{cc}(s) \right) \mathbf{i} & = G_\text{cc}(s) Y_\text{v}(s) \left( \mathbf{e} - \mathbf{v} \right) 
    \\
    \Rightarrow
    \mathbf{v} & = \mathbf{e} - \frac{sL_\text{f} + G_\text{cc}(s)}{G_\text{cc}(s) Y_\text{v}(s)} \mathbf{i},
  \end{aligned}
  \label{eq:Simple_V_2}
\end{equation}
reflecting the overall interaction among VA, CC, and VFF.
Equation~\eqref{eq:Simple_V_2} implies that VA-CC emulates the impedance between
the PCC and the IVS as \( \left( sL_\text{f} + G_\text{cc}(s) \right) / \left( G_\text{cc}(s) Y_\text{v}(s) \right) \).
Note that, for \(\omega \rightarrow \omega_1\), \( \left| G_\text{cc}(j\omega) \right| \gg K_\text{p}\)
due to the resonant controller. Therefore, VA-CC can accurately emulate the
targeted impedance \(1/Y_\text{v}(j\omega_1)\) at the
fundamental frequency. However, the resonant controller
is frequency-selective;
it applies only to specific frequency components and should not interfere with other controllers.
Consequently, the gain of the CC cannot be infinite across the entire frequency range.
Rather, it is commonly assumed that \(G_\text{cc}(s) \approx K_\text{p} = \omega_\text{cc}L_\text{f}\) in the harmonic range~\cite{Harnefors2015Passivity}.

\begin{figure}[!t]
    \centering
    \subfloat[]{
        \includegraphics[width=0.55\linewidth]{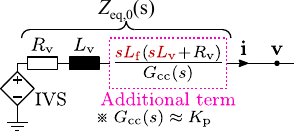}
    }
    \vspace{0pt}
    \subfloat[]{
        \includegraphics[width=0.625\linewidth]{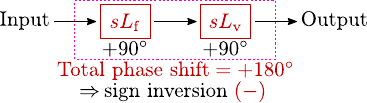}
    }  
    \caption{Negative-resistance caused by intra-loop coupling within VA-CC:
    (a) equivalent circuit of output impedance \(Z_\text{eq,0}(s)\), (b) emergence of a control-delay-independent negative-resistance term, \(s^2\).}
    \label{fig:Diagram_s2}
\end{figure}

In the harmonic range, where \(K_\text{p}\) is the dominant term of \( G_\text{cc} (s) \), the term \(sL_\text{f}\) is no longer negligible---especially
considering that \( \left| j\omega L_\text{f} \right| \rightarrow \infty \) as \(\omega \rightarrow \infty \).
Therefore, the equivalent output impedance of the inverter is derived as:
\begin{equation}
  \begin{aligned}
    & Z_\text{eq,0} (s) = \frac{sL_\text{f}}{G_\text{cc} (s) Y_\text{v}(s)} + \frac{1}{Y_\text{v}(s)}
     = \underbrace{\frac{s L_\text{f} \left( sL_\text{v} + R_\text{v} \right) }{G_\text{cc}(s)}}_{s^2\text{-term included}} + \frac{1}{Y_\text{v}(s)}
    \\ & \approx \underbrace{\frac{s L_\text{f} \left( sL_\text{v} + R_\text{v} \right) }{K_\text{p} }}_{s^2\text{-term included}} + \frac{1}{Y_\text{v}(s)}
    = \underbrace{{s^2} \frac{L_\text{v}L_\text{f}}{K_\text{p}} }_{s^2\text{-term}}+ s \frac{ R_\text{v}}{\omega_\text{cc}} + s L_\text{v} + R_\text{v}
  \end{aligned}
\label{eq:Simple_Zeq}
\end{equation}
where \(Z_\text{eq,0}(s)\) denotes the delay-absent equivalent output impedance of VA-CC.
Due to the limitation of \(\omega_\text{cc}\),
as depicted in Fig.~\ref{fig:Diagram_s2}(a), additional unwanted terms besides the virtual impedance arise:
\begin{equation}
  \underbrace{s^2 \frac{L_\text{v}L_\text{f}}{K_\text{p}}}_{s = j\omega\text{: negative real}} + s \frac{R_\text{v}}{\omega_\text{cc}},
\end{equation}
which result from the intra-loop coupling within VA-CC.
Notice that the term \(s^2\) appears explicitly in \(Z_\text{eq,0}(s)\).
This term arises from the product of two inductive terms,
\(sL_\text{f}\) and \(sL_\text{v}\),
as demonstrated in Fig.~\ref{fig:Diagram_s2}(b).
An important property of the \(s^2\)-term is that
the phase response of \(s^2\) is \(180^\circ\) for \(s = j\omega\).

Consider \(s = j\omega\). Then the real part of \(Z_\text{eq,0}(j \omega)\) becomes
\begin{equation}
  \Re \left[ Z_\text{eq,0}(j\omega) \right] = - \omega^2 L_\text{v}/\omega_\text{cc} + R_\text{v},
  \label{eq:NonPassive_Simple}
\end{equation}
with \(K_\text{p} =\omega_\text{cc}L_\text{f}\).
It is immediately apparent that \( - \omega^2 L_\text{v}/\omega_\text{cc} \)
inevitably induces a negative-resistance part in the equivalent output impedance.
Note that the \(s^2\)-term scales quadratically with frequency,
thus becoming dominant in the harmonic range.
Accordingly, for the frequency range
\begin{equation}
\omega > \sqrt{\omega_\text{cc}R_\text{v}/L_\text{v}} =   \sqrt{\omega_\text{cc}\omega_\text{1}/n_\text{XR}},
\end{equation}
\(\Re \left[ Z_\text{eq,0}(j\omega) \right] < 0\).
Thus, the \(s^2\)-term induces a negative-real part, causing negative-resistance property
regardless of the control delay.
Due to the \(s^2\)-term, the critical frequency \(f_\text{cr}\) that
determines the passivity range of VA-CC is therefore
\begin{equation}
  f_\text{cr} = \frac{1}{2 \pi} \sqrt{\frac{\omega_\text{cc} \omega_1}{n_\text{XR}}}
  \label{eq:f_cr}
\end{equation}
in Hz. Above this frequency, the VA-CC is likely to induce a negative-resistance property.

\begin{figure}[!t]
    \centering
    \subfloat[]{
        \includegraphics[width=0.9\linewidth]{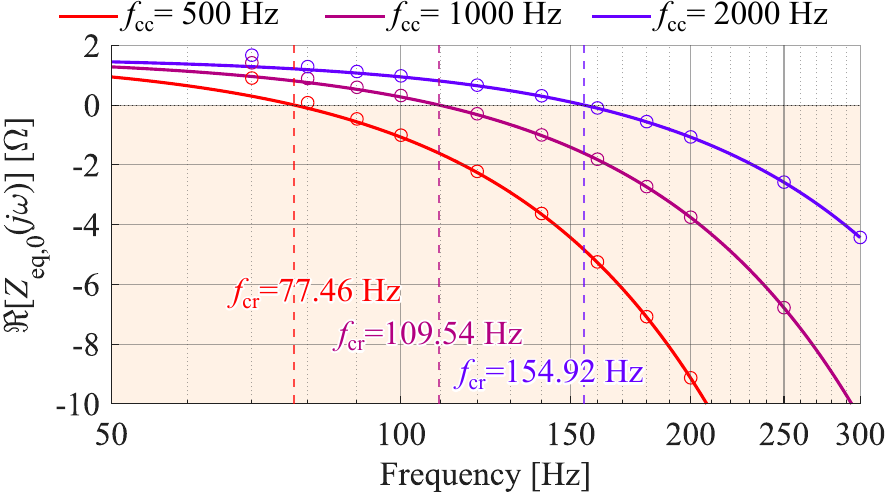}
        \label{fig:Validation1_BW}
    }
    \vspace{0pt}
    \subfloat[]{
        \includegraphics[width=0.9\linewidth]{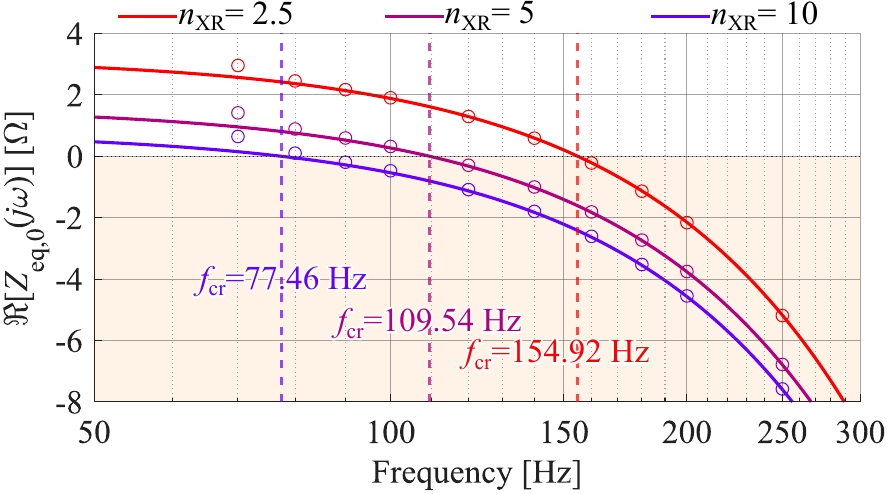}
        \label{fig:Validation1_XR}
    }  
    \caption{Real part of \(Z_\text{eq,0}(j\omega)\) in the frequency range up to 300~Hz,
    (a) under different \(f_\text{cc}\) and fixed \(n_\text{XR} = 5\),
    (b) under different \(n_\text{XR}\) and fixed \(f_\text{cc} = 1000\)~Hz.
    The circle plots represent the simulation-scanned results.
    }
    \label{fig:Validation1}
\end{figure}

To further verify the identified negative-resistance mechanism,
and to examine whether the effects of the outer loop and the resonant controller are negligible,
the real part of \(Z_\text{eq,0}(s)\) is displayed in the relatively low harmonic
range up to 300~Hz in Fig.~\ref{fig:Validation1}.
The analytical results are compared with the time-domain simulation-scanned results.
The simulation-scanned results reflect the delay-free model including
the CC with PR control and the outer loop.

In both Fig.~\ref{fig:Validation1}(a) and (b), \(L_\text{v}\) is set to 0.5~p.u.,
which should ensure low-frequency stability of the GFM inverter under a strong grid condition \cite{Zhao2024Low}.
The real part of \(Z_\text{eq,0}(s)\) is displayed
in Fig.~\ref{fig:Validation1}(a) and Fig.~\ref{fig:Validation1}(b).
Fig.~\ref{fig:Validation1}(a) presents cases with various CC bandwidths and a fixed \(n_\text{XR} = 5\), and
Fig.~\ref{fig:Validation1}(b) presents cases with various \(n_\text{XR}\) and a fixed \(f_\text{cc} = 1\)~kHz.
The remaining parameters are detailed in Table~\ref{tab:SectionII}.
In all cases, the equivalent output impedance readily falls into the negative-real region.
That is, it exhibits negative-resistance property.
The predicted \(f_\text{cr}\) from the simplified model in
\eqref{eq:f_cr} is also displayed in Fig.~\ref{fig:Validation1}(a) and (b),
and aligns the simulation-scanned result in all cases.
This match demonstrates that the \(s^2\)-term,
induced by the control-loop interaction within VA-CC itself,
is the cause of the negative-resistance property.
The difference between \eqref{eq:Simple_Zeq} and the simulation-scanned result
appears only near the fundamental frequency,
which is due to the impact of the outer loop and the resonant controller.
Nevertheless, this impact is limited to the region
near the fundamental frequency.


\begin{figure}[!t]
    \centering
    \subfloat[]{
        \includegraphics[width=0.88\linewidth]{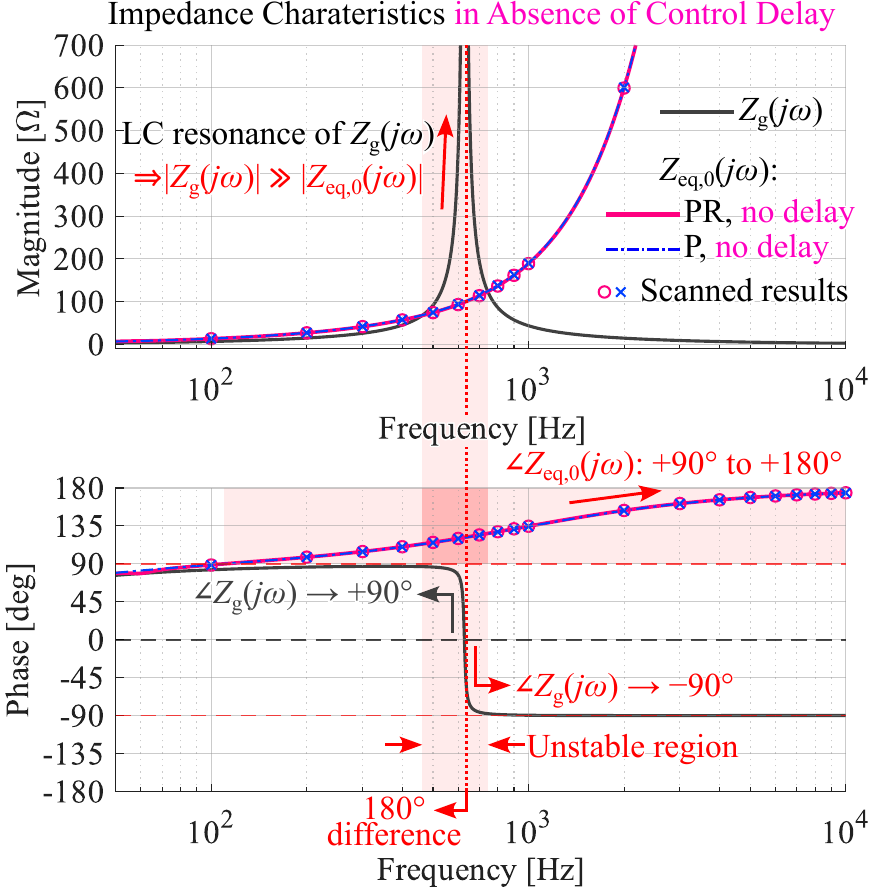}
    }
    \vspace{0pt}
    \subfloat[]{
        \includegraphics[width=0.88\linewidth]{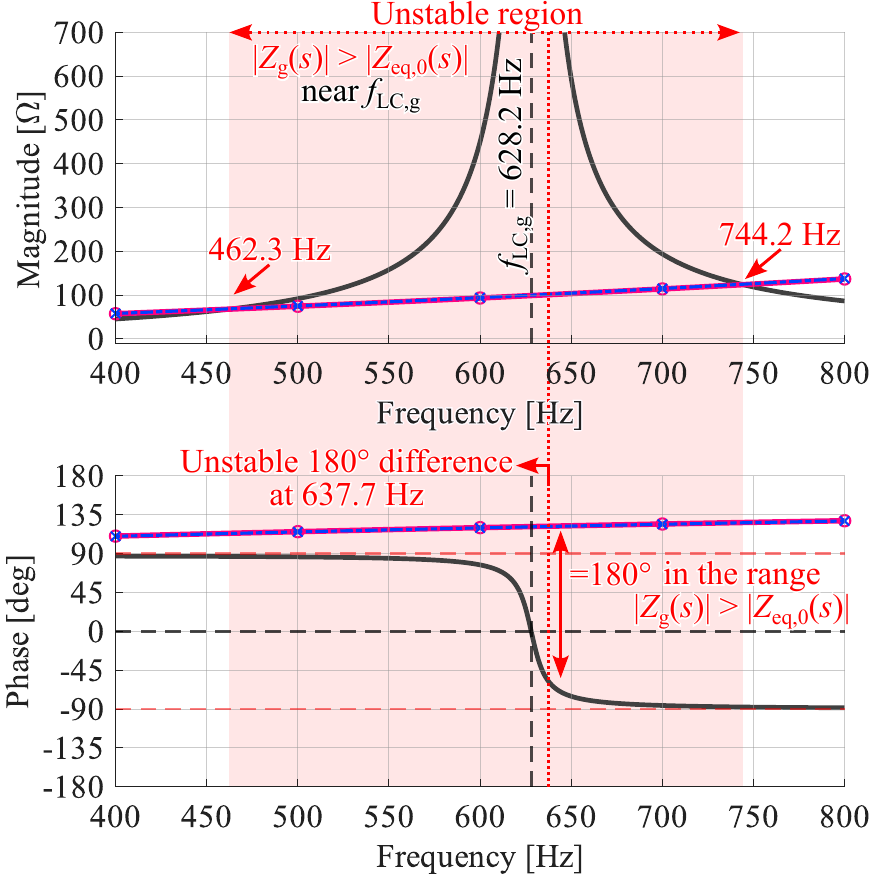}
    }  
    \caption{Impedance plots of \(Z_\text{eq,0}(s)\) and \(Z_\text{g}(s)\).
    (a) Plots of \(Z_\text{eq,0}(s)\) and \(Z_\text{g}(s)\),
    (b) magnified view from 400 to 800~Hz.
    The circle plots represent the simulation-scanned results.
    }
    \label{fig:Validation2_NoDelayPlot}
\end{figure}

\subsection{Analysis of Harmonic Instability in VA-CC in the Absence of Control Delay}

From the previous analysis, \(Z_\text{eq,0} (s)\) is shown to exhibit
second-order dynamics. As it takes the form \( m_2 s^2 + m_1 s + m_0\),
the phase \(\angle Z_\text{eq,0}(j \omega)\) ranges from 0 to 180 degrees as \(\omega\) varies from 0 to \(\infty\).
Therefore, the \(180^\circ\) phase difference does not occur
if the grid-side impedance \(Z_\text{g}(s)\) is purely inductive over the entire frequency range,
i.e., \(\angle Z_\text{g}(j\omega) \in \left(0^\circ, 90^\circ \right)\) for any \(\omega\).
However, in practice,
shunt capacitance is typically present.
This may arise from an LC-filter capacitor of the inverter itself,
a capacitor bank of the grid,
or cable capacitance in the case of an offshore wind farm~\cite{Cespedes2014Mitigation}.
Due to the shunt capacitor, an LC resonant frequency exists in the grid-side impedance,
which induces the following conditions under which the VA-CC may produce
harmonic instability:
\begin{enumerate}
  \item At the grid-side LC resonant frequency, $ f_\text{LC,g} $, the grid impedance exhibits a large magnitude,
  $ |Z_\text{g} (s)| > |Z_\text{eq,0} (s)| $.
  \item Simultaneously, the phase $ \angle Z_\text{g} (s) $ flips from inductive to capacitive.
  If $ \angle Z_\text{eq,0} (s) \in (90^\circ, 180^\circ) $, there exists a frequency at which the $ 180^\circ $ phase difference occurs between
  $ \angle Z_\text{eq,0} (s) $ and $ \angle Z_\text{g} (s) $.
\end{enumerate}
Due to these two simultaneous events, an unstable \(180^\circ\) phase difference is created
between the grid impedance and the inverter's equivalent output impedance near the
grid-side LC resonant frequency, \(f_\text{LC,g}\).

The impedance plot---including two cases of the CC with P and PR control, the simulation-scanned results,
and the grid impedance---is shown in Fig.~\ref{fig:Validation2_NoDelayPlot}.
As discussed before, the effect of R controller in CC on the impedance characteristic is negligible,
and the results of the developed impedance model coincide with the scanned results precisely.
The parameters are summarized as Table~\ref{tab:SectionII}.
According to the analysis,
the impedance plot shown in Fig.~\ref{fig:Validation2_NoDelayPlot}
demonstrates that the LC resonance magnitude is significantly larger than the magnitude of \(Z_\text{eq,0}(s)\)
at the LC resonant frequency.
Meanwhile, \(\angle Z_\text{eq,0}(s)\) lies between \(+90^\circ\) to \(+180^\circ\) due to the \(s^2\)-term.
Therefore, a region with a negative stability margin is induced
between 462.3~Hz and 744.2~Hz.
This aligns with the simulation result shown in Fig.~\ref{fig:Validation2_NoDelaySimulation},
which shows the main oscillation at approximately 714~Hz, lying in the identified unstable
frequency range.


\section{Detailed Analysis of Negative-Resistance Property in Inverters with Conventional VA-CC}
\label{Section:III}

In physical implementation, a control delay is inevitable and must be considered
in an elaborate manner.
Therefore, a detailed model using first-order Pad\'e approximation
of the control delay is considered \cite{Liu2021Small, Yang2005Applied}.
Using this model, the frequency range over which
the delay-considered impedance model exhibits the
negative-resistance property is investigated.


\subsection{Detailed Equivalent Output Impedance of VA-CC Considering Control Delay}

\begin{figure}[!t]
    \centering
    {
        \includegraphics[width=0.925\linewidth]{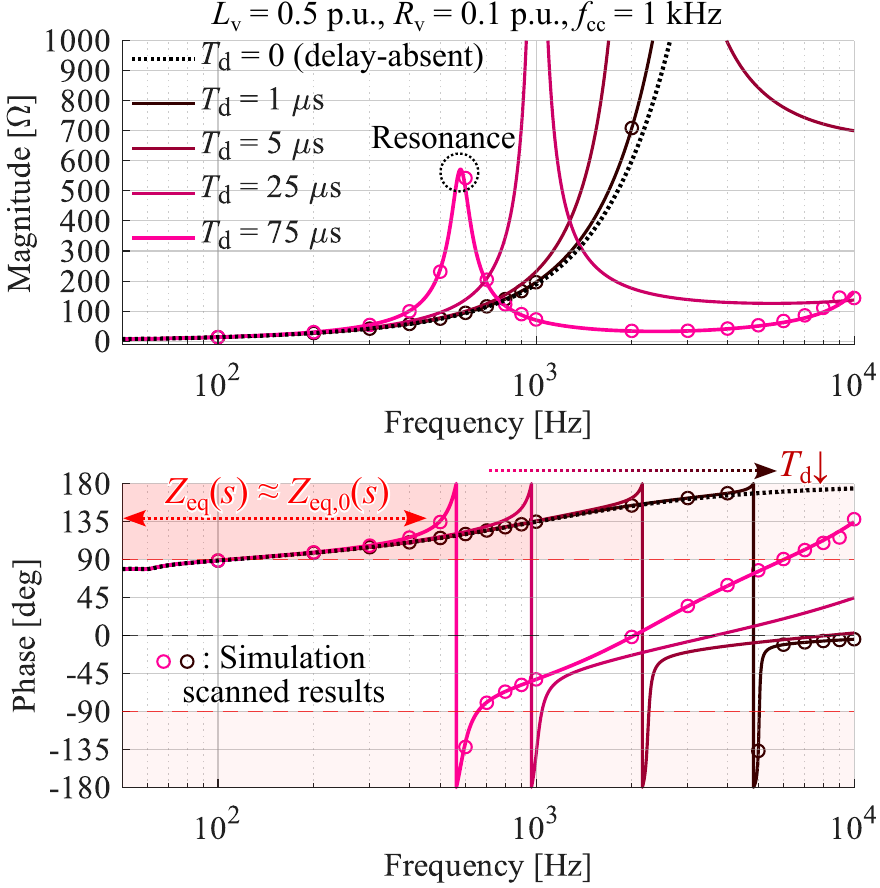} 
    }
    \caption{Comparison of \(Z_\text{eq}(s)\) and \(Z_\text{eq,0}(s)\),
    together with the scanned results of the time-domain simulation,
    for different amounts of control delay \(T_\text{d}\).
    Control parameters are
    \(f_\text{cc} = 1\)~kHz, \(L_\text{v} = 0.5\)~p.u., and \(n_\text{XR} = 5\).}
    \label{fig:Validate3_Approach}
\end{figure}


With the control delay taken into account, the modulated output voltage can be written as
\begin{equation}
  \mathbf{v}_\text{inv} = G_\text{d}(s) G_\text{cc} (s) Y_\text{v}(s) \left( \mathbf{e} - \mathbf{v} \right)
  - G_\text{d}(s) G_\text{cc}(s) \mathbf{i} + G_\text{d}(s)  \mathbf{v}.
\label{eq:Full_Vinv}
\end{equation}
The full model of the equivalent output impedance, \(Z_\text{eq}(s)\), is expressed as:
\begin{equation}
  \begin{aligned}
  Z_\text{eq}(s) & = - \frac{ \Delta \mathbf{v} }{ \Delta \mathbf{i} }
  = \frac{ G_\text{d}(s) G_\text{cc}(s) + sL_\text{f} }{1 - G_\text{d}(s) + G_\text{d}(s) G_\text{cc}(s) Y_\text{v}(s)}.
  \end{aligned}
  \label{eq:Full_Zeq}
\end{equation}
where \(\Delta \mathbf{e} = 0\) is assumed, owing to the slow IVS dynamics.

To verify the consistency with the delay-absent model \(Z_\text{eq,0}(s)\)
developed in the previous section,
the full model \(Z_\text{eq}(s)\) is plotted for several values of control delay
(1, 5, 25, and 75~\(\mu\)s) 
in Fig.~\ref{fig:Validate3_Approach}.
The scanned results of the time-domain simulation for the minimum and
maximum \(T_\text{d}\) (1 and 75~\(\mu\)s) are also plotted for verification.
As expected, \(Z_\text{eq}(s)\) converges to \(Z_\text{eq,0}(s)\)
as the control delay approaches zero.
When the control delay is taken into account,
unlike in the delay-absent condition,
an inverter resonant frequency appears.
Up to around this inverter resonant frequency, \(Z_\text{eq}(s)\)
exhibits a negative-resistance property,
consistent with the delay-absent condition.
This can be interpreted as follows:
even when the control delay is taken into account,
the negative-resistance mechanism discussed in Section~\ref{Section:II} still emerges
in the frequency range where the effect of control delay is less significant.
The location of the resonant frequency,
which determines the range of the negative-resistance property,
is further examined in the next subsection.

\subsection{Detailed Negative-Resistance Frequency Range in VA-CC}
\label{Subsection:III-B}

Assuming \(|sT_\text{d}| < 1\), the first-order Pad\'e approximation of the control delay
\(G_\text{d}(s) = e^{-sT_\text{d}}\) is 
\(e^{-sT_\text{d}} \approx \frac{1 - s T_\text{d}/2}{1 + s T_\text{d}/2}\).
Substituting this and \eqref{eq:Simple_V} into \eqref{eq:Full_Zeq}
results following approximation, \(Z_\text{eq,Pade}(s)\), assuming \(G_\text{cc}(s) \approx K_\text{p}\):
\begin{equation}
  \begin{aligned}
  Z_\text{eq}(s) & = - \frac{\Delta \mathbf{v} }{\Delta \mathbf{i} }
  \approx \frac{ \frac{1 - s T_\text{d} / 2}{1 + s T_\text{d} / 2} K_\text{p} + sL_\text{f} }
  { 1 - \frac{1 - s T_\text{d} / 2}{1 + s T_\text{d} / 2} + \frac{1 - s T_\text{d} / 2}{1 + s T_\text{d} / 2} K_\text{p} Y_\text{v}(s)}
  \\ &
  = \frac{\left( sL_\text{v} + R_\text{v} \right) N(s)}{D(s)}
  = Z_\text{eq,Pade}(s)
    \end{aligned}
  \label{eq:Full_Zeq_approx}
\end{equation}
where 
\begin{equation}
  \begin{aligned}
    N(s) = & \,\ s^2 \frac {T_\text{d} L_\text{f}}{2} + 
    s \left( L_\text{f} -  \frac{T_\text{d} K_\text{p}}{2} \right)
    + K_\text{p},
  \end{aligned}
  \label{eq:Numerator}
\end{equation}
\begin{equation}
  \begin{aligned}
    D(s) = & \,\ s^2 T_\text{d}L_\text{v}
    + s \left( T_\text{d} R_\text{v} - \frac{T_\text{d} K_\text{p}}{2} \right)
    + K_\text{p}.
  \end{aligned}
  \label{eq:Denominator}
\end{equation}

Under the condition where \(sT_\text{d} \approx 0 \),
\eqref{eq:Numerator} and \eqref{eq:Denominator} can be approximated
as \( \left( sL_\text{f} + K_\text{p} \right) \) and \(K_\text{p}\), respectively.
In this case, the output impedance of \eqref{eq:Full_Zeq_approx} reduces to \eqref{eq:Simple_Zeq}.
Therefore, it is expected that the negative-resistance property,
induced by the intra-loop coupling,
appears in the relatively low harmonic range where the control delay can be neglected.

Considering the control delay, the natural frequencies of \(N(s)\) and \(D(s)\) are
\begin{equation}
  \omega_N = \sqrt{ \frac{2 K_\text{p}}{T_\text{d} L_\text{f}} }
   = \sqrt{ \frac{2 \omega_\text{cc}}{T_\text{d}} },
   \label{eq:omega_N}
\end{equation}
\begin{equation}
  \omega_D = 2 \pi f_{D} = \sqrt{ \frac{K_\text{p}}{T_\text{d} L_\text{v}} }
   = \sqrt{ \frac{\omega_\text{cc} L_\text{f}}{T_\text{d} L_\text{v}} }.
   \label{eq:omega_D}
\end{equation}
Under practical parameter design,
two observations follow: (a) \(\omega_D < \omega_N\), and
(b) the denominator \(D(s)\) is typically underdamped, i.e., its damping coefficient \(\zeta_D\) is small.
The detailed reasoning for (a) and (b)
is provided in the Appendix.
Note that, under these conditions,
\(\angle N(s)\) lies between \(0^\circ\) to \(90^\circ\)
for \(\omega < \omega_N\).
Therefore, for \(s=j\omega\) such that \(0<\omega < \omega_D < \omega_N\),
\(\angle N(s)\) remains inductive.
Meanwhile, \(\angle D(s) \approx 0^\circ\) below \(\omega_D = 2 \pi f_D\),
whereas it quickly approaches \(\pm180^\circ\) above \(\omega_D\).

\begin{figure}[!t]
    \centering
    \subfloat[]{
        \includegraphics[width=0.95\linewidth]{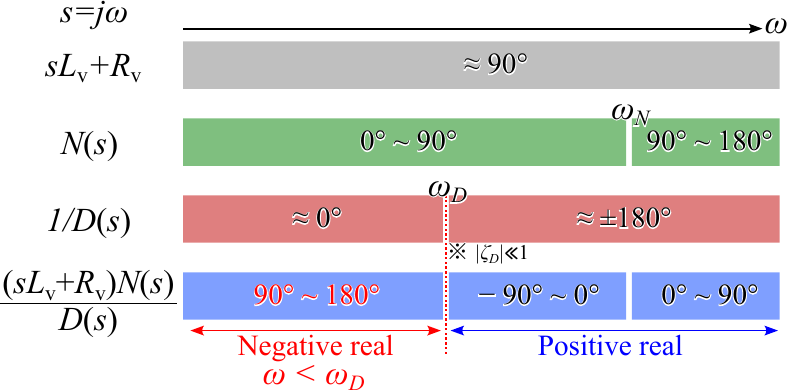}
        \label{fig:Validate3_Diagram}
    }
    \vspace{0pt}
    \subfloat[]{
        \includegraphics[width=0.75\linewidth]{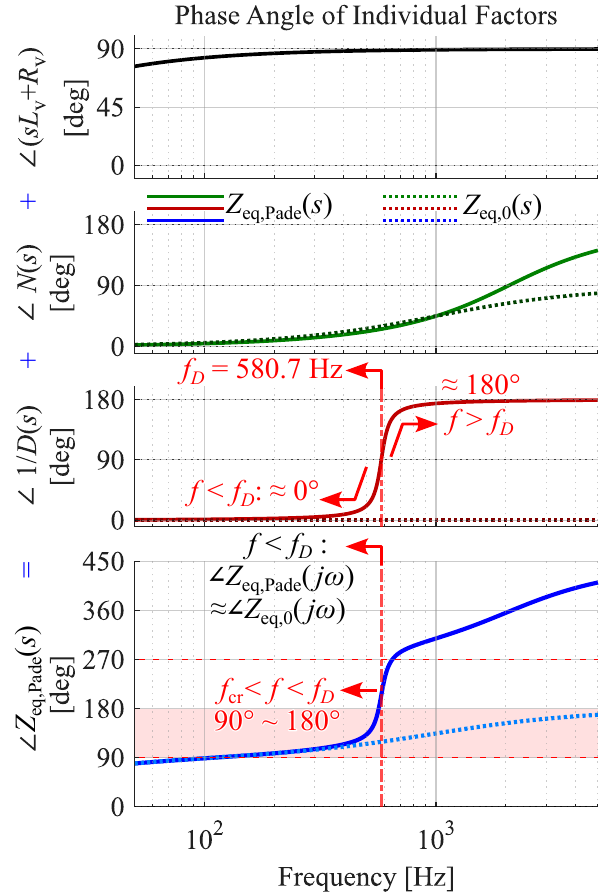}
        \label{fig:Validate3_Phase}
    }
    \caption{Diagram of phase angle responses of each approximated factors.
    (a) Schematic illustration.
    (b) Actual phase contribution of the individual factors,
    with $ f_\text{cc} = 1 $~kHz, and the remaining parameters given in Table~\ref{tab:SectionIII}.}
    \label{fig:Validation3_Diagram}
\end{figure}

\begin{table}[t]
\centering
\caption{Section III, IV, V Parameters: Default Values}
\label{tab:SectionIII}
\begin{tabular}{c c c}
\hline
\text{Parameter} & \text{Symbol} & \text{Value} \\
\hline
Fundamental grid frequency & \(f_1\) (\(\omega_1\)) & 60~Hz (\(120\pi\)~rad/s)\\
Switching frequency & \(f_{\text{sw}}\) & 10~kHz \\
Sampling frequency & \(f_{\text{s}}\) & 20~kHz \\
Control delay & \(T_{\text{d}}\) & 75~\(\mu\)s\(^{[*]}\) \\
Rated apparent power & \(S_{\text{n}}\) & 3.0 kVAR \\
DC-link voltage & \(V_{\text{dc}}\) & 400 V \\
Grid voltage, line-to-line RMS & \(V_{\text{g,rms}}\) & 220 V \\
Grid short circuit ratio & SCR & 4.0 \\
Grid X/R ratio & \(n_\text{XR,g}\) & 5 \\
Damping resistance & {\(R_{\text{c}}\)} & {10 m\(\Omega\)} (6.2e-4~p.u.) \\
Filter inductance & {\(L_{\text{f}}\)} & {3.4 mH} (0.079~p.u.) \\
Cut-off frequency of the outer loop & {\(\omega_{{P}}, \omega_{{Q}}\)} & \(6\pi\)~rad/s (0.05~p.u.) \\
Droop gains of the outer loop & {\(D_{{P}}, D_{{Q}}\)} & 0.05~p.u. \\
Virtual inductance & {\(L_{\text{v}}\)} & {21.4 mH} (0.5~p.u.)\\
X/R ratio of conventional VA & {\(n_{\text{XR}}\)} & 5 \\
\hline
\end{tabular}
\begin{tablenotes}
\small
\item{[*] \(T_\text{d} = 1.5/f_\text{s}\).}
\end{tablenotes}
\end{table}

To better understand the phase response of \(Z_\text{eq,Pade}(s)\),
a diagram demonstrating the phase contribution of each element is given in
Fig.~\ref{fig:Validation3_Diagram}(a) and (b).
Fig.~\ref{fig:Validation3_Diagram}(a) shows the contributions schematically,
whereas
Fig.~\ref{fig:Validation3_Diagram}(b)
illustrates the impact of each factor on \(\angle Z_\text{eq,Pade}(s)\) for the specific case
\(f_\text{cc} = 0.05 f_\text{s} = 1\)~kHz.
The remaining parameters are given in Table~\ref{tab:SectionIII}.
Fig.~\ref{fig:Validation3_Diagram}(b) plots the
phase responses of the individual factors of \(Z_\text{eq,Pade}(s)\) (solid lines)
and \(Z_\text{eq,0}(s)\) (dotted lines) for comparison.
\(\angle Z_\text{eq,Pade}(s)\) is decomposed as the sum of
\(\angle \left( j \omega L_\text{v} + R_\text{v} \right) \approx 90^\circ \),
\(\angle N(s)\) 
and \( \angle 1/D(s) \).
Below \(f_D\), \( \angle 1/D(s) \approx 0^\circ \), so the total phase lies in
\( \left[ 90^\circ, 180^\circ \right]\), corresponding to the aforementioned
\(s^2\)-term-induced negative-resistance property. 
Above \(f_D\), \( \angle 1/D(s)\) rapidly shifts away from \(0^\circ \),
and the total phase deviates from that of \(\angle Z_\text{eq,0}(s)\).
Therefore, the negative-resistance property of
the delay-considered model persists up to approximately \(\omega_D = 2 \pi f_D\)
determined by \eqref{eq:omega_D}.

\begin{figure}[!t]
    \centering
        \includegraphics[width=0.9\linewidth]{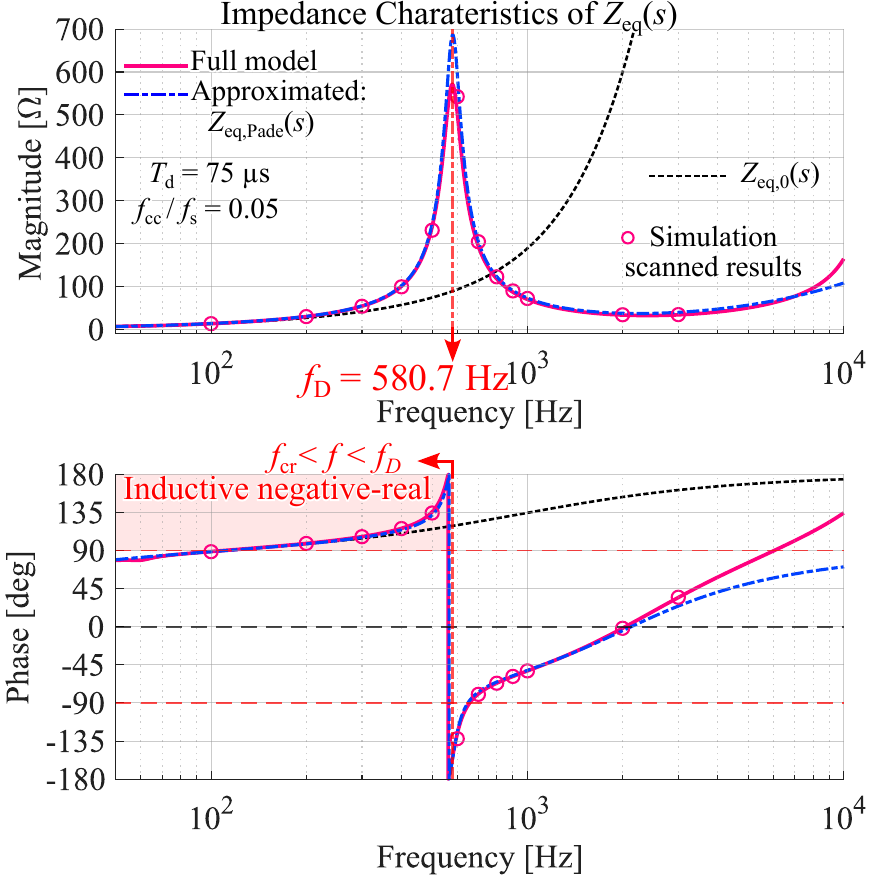}
    \caption{Comparison of the full model \(Z_\text{eq}(s)\),
    the approximated model \(Z_\text{eq,Pade}(s)\),
    and the delay-absent model \(Z_\text{eq,0}(s)\), together with the
    scanned results of the time-domain simulation, for 
    \(f_\text{cc} = 1\)~kHz.
    }
    \label{fig:Validate3_Impedance plot}
\end{figure}

For verification of the approximation, impedance plots are compared and shown in Fig.~\ref{fig:Validate3_Impedance plot}.
The full model includes the PR control and the control delay modeled fully as \(e^{-sT_\text{d}}\).
This is compared with the approximated model \(Z_\text{eq,Pade}(s)\),
along with the scanned results from the time-domain simulations.
The result shows that the proposed approximation is valid over a wide harmonic range.
The CC bandwidth is set to \(f_\text{cc} = 0.05 f_\text{s} = 1\)~kHz.
In the plot, the value of \(f_D\) is analytically determined by \eqref{eq:omega_D},
which closely matches the resonant peak of the actual full model.
Near and below \(f_D\), \(\angle Z_\text{eq}(s)\)
exhibits a negative-resistance property.

The following subsections demonstrate that 
the VA-CC is likely to trigger harmonic instability,
particularly when the grid-side LC resonant frequency
is lower than \(f_D\).
This results from the interaction between
the aforementioned negative-resistance property
and the grid-side LC resonance,
which is the same harmonic instability mechanism shown in Fig.~\ref{fig:Validation2_NoDelaySimulation}
and Fig.~\ref{fig:Validation2_NoDelayPlot}.

\subsection{Stability of VA-CC under Different CC Bandwidths}
\label{Subsection:III-C}

\begin{figure}[!t]
    \centering
    {
        \includegraphics[width=0.975\linewidth]{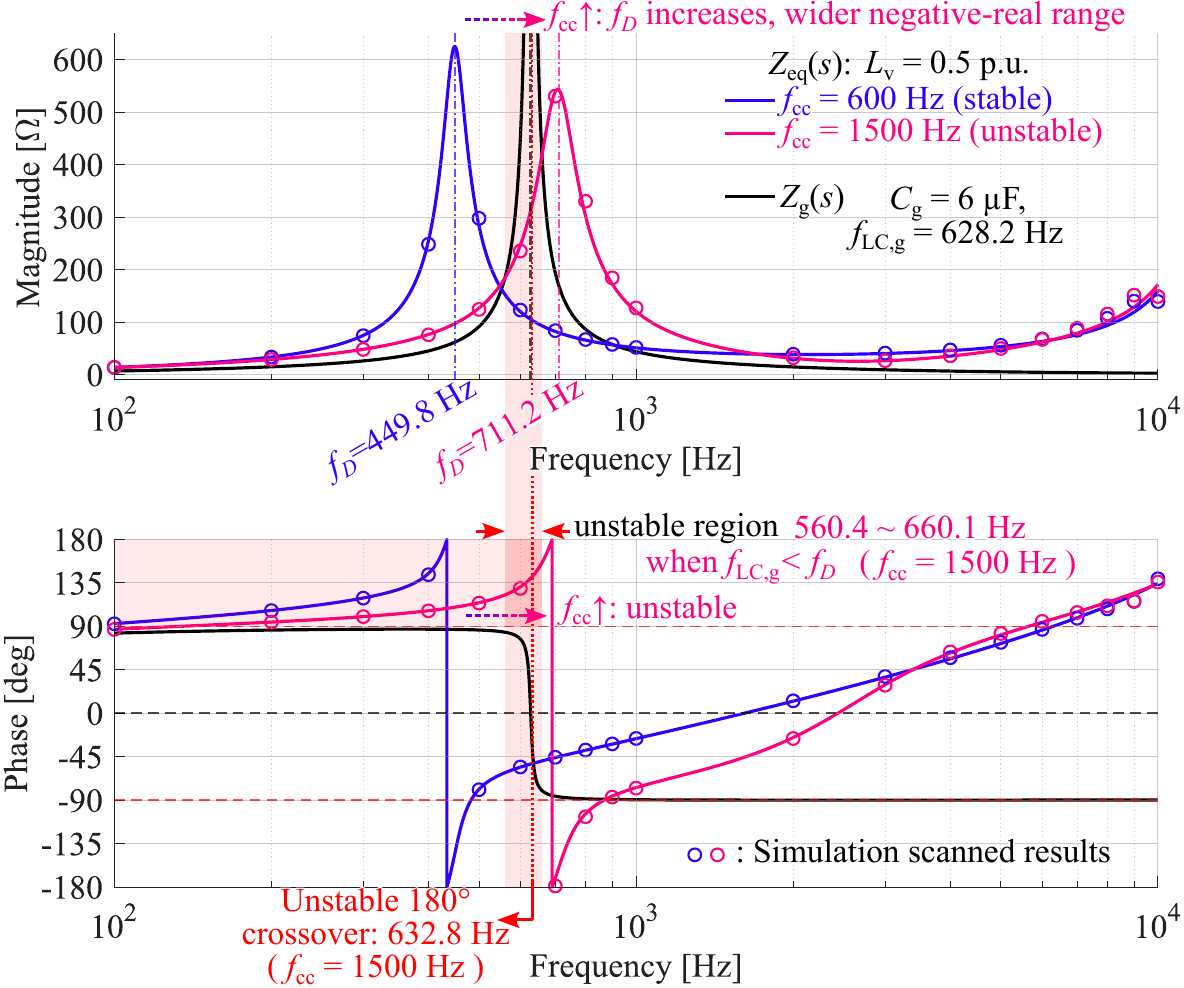}
    }
    \caption{Plot of \(Z_\text{g}(s)\) and \(Z_\text{eq}(j\omega)\) for two different CC bandwidths:
    $f_\text{cc}= 0.03f_\text{s} =600$~Hz (stable) and $f_\text{cc} = 0.075f_\text{s} = 1.5$~kHz (unstable).
    }
    \label{fig:Validation4_CC}
\end{figure}

Fig.~\ref{fig:Validation4_CC} demonstrates the  cases
under two different CC bandwidths, 600~Hz and 1.5~kHz.
The grid-side shunt capacitance \(C_\text{g}\) is fixed at 6~\(\mu\)F.
The remaining parameters are detailed in Table~\ref{tab:SectionIII}.
Compared to the sampling frequency of 20~kHz, the two CC bandwidths corresponds to
0.03\(f_\text{s}\) (600~Hz) and 0.075\(f_\text{s}\) (1.5~kHz).
For a higher CC bandwidth, \(f_D\) increases according to \eqref{eq:omega_D},
and therefore the negative-resistance region widens.
In the cases of 600~Hz and 1.5~kHz,
\(f_D\) is 449.8~Hz and 711.2~Hz, respectively.

For the case with a higher CC bandwidth of \(f_\text{cc} = 1.5\)~kHz,
\(Z_\text{eq}(s)\) exhibits a negative-resistance property
over a wide frequency range, and \(f_\text{LC,g}\) lies within this range.
As in the case shown in Fig.~\ref{fig:Validation2_NoDelayPlot},
an unstable frequency region also appears in Fig.~\ref{fig:Validation4_CC}.
Due to the interaction between the grid-side LC resonance and the
negative resistance,
a \(180^\circ\) phase difference
emerges at 632.8~Hz,
in the range where \(|Z_\text{g}(s)| > |Z_\text{eq}(s)|\).
Therefore, the unstable frequency range is determined to be
560.4~Hz to 660.1~Hz.

Since \(f_\text{cc}\) is typically employed up to \(0.1f_\text{s}\) \cite{Harnefors2016Passivity},
conventional VA-CC does not fully leverage the benefit of
the high sampling frequency and low control delay of \(f_\text{s} = 20\)~kHz and \(T_\text{d} = 75~\mu\)s.
Rather, \(f_\text{cc}\) must be limited owing to the
control-delay-independent non-passivity.


\begin{figure}[!t]
    \vspace{5pt}
    \centering
    {
        \includegraphics[width=0.975\linewidth]{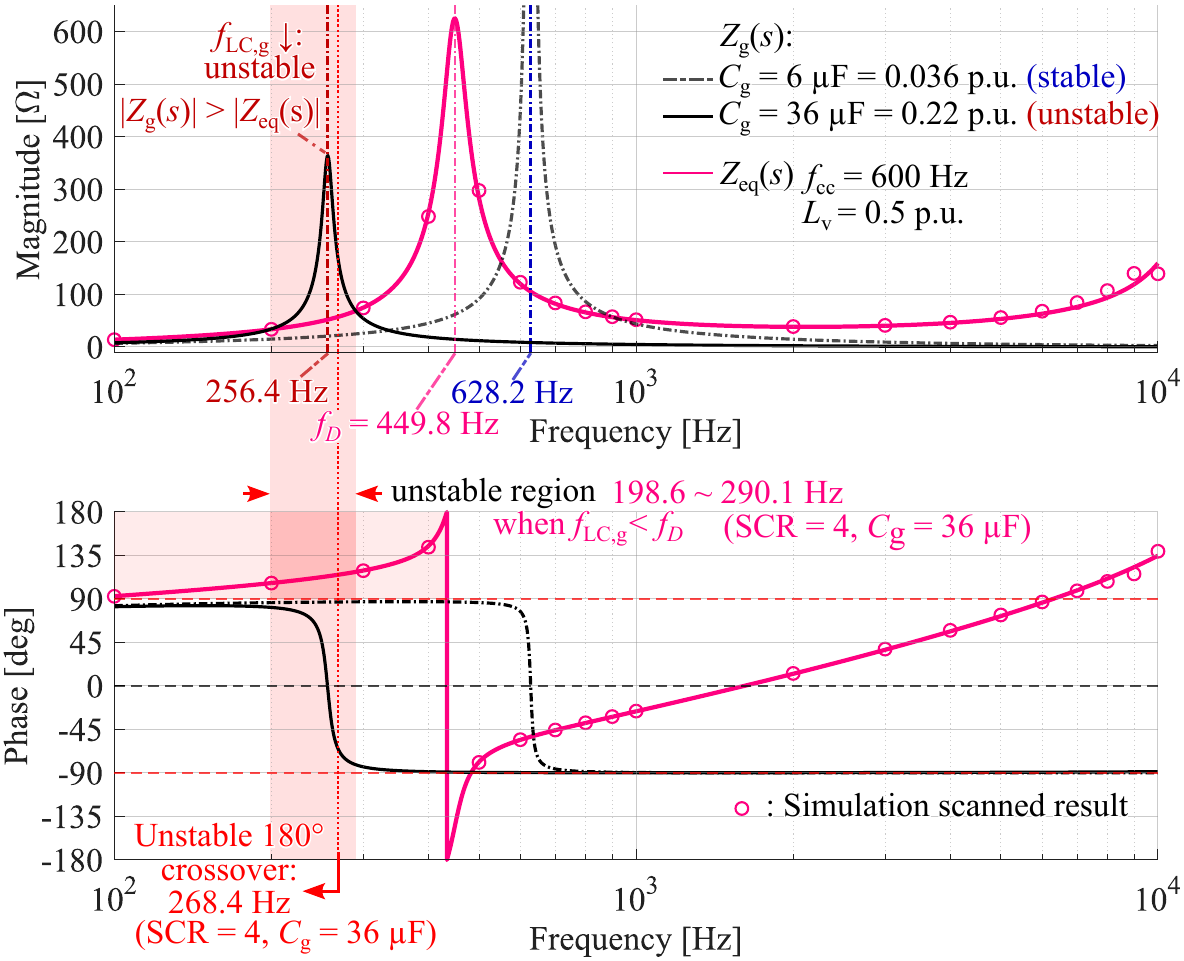}
    }
    \caption{Comparison of \(Z_\text{eq}(s)\) and two different \(Z_\text{g}(s)\) for two different \(C_\text{g}\):
    $C_\text{g}=6~\mu$F (stable), $C_\text{g}=36~\mu$F (unstable).
    }
    \label{fig:Validation4_Grid}
\end{figure}

\subsection{Stability of VA-CC under Different Grid Conditions}
\label{Subsection:III-D}

One might attempt to mitigate
the instability of VA-CC by adjusting \(f_\text{cc}\)
to avoid harmonic instability under a given grid condition.
Specifically, \(f_D\) can be reduced by decreasing
\(f_\text{cc}\), intentionally limiting
the CC performance despite the high sampling frequency and low control delay.
However, this approach may still fail when the grid condition changes,
especially when \(f_\text{LC,g}\) decreases.

Consider a scenario in which \(f_\text{cc}\) is lowered to 600~Hz (0.03\(f_\text{s}\)).
The remaining parameters are detailed in Table~\ref{tab:SectionIII}.
Two grid conditions are considered in Fig.~\ref{fig:Validation4_Grid},
corresponding to different grid-side shunt capacitances,
with \(C_\text{g}\) of 6~\(\mu\)F and 36~\(\mu\)F.
In each case, \(f_\text{LC,g}\) corresponds to
628.2~Hz and 256.4~Hz, respectively.
When \(f_\text{LC,g}\) is considerably larger than
\(f_{D} = 449.8\)~Hz, the system remains stable,
since the grid-side LC resonance does not
occur in the negative-resistance region of the VA-CC-based inverter.
However, when \(f_\text{LC,g}\) becomes smaller, i.e., for \(C_\text{g}=36~\mu\)F,
\(f_\text{LC,g}\) can fall below
\(f_{D} = 449.8\)~Hz, and the system can become unstable.
In this case with \(f_\text{LC,g} = 256.4\)~Hz,
the grid-side LC resonance peak occurs in the frequency region
where \(Z_\text{eq}(s)\) exhibits the negative-resistance property.
Therefore, an unstable \(180^\circ\) phase difference occurs
at 268.4~Hz, and the unstable frequency range is 198.6~Hz to 290.1~Hz.

This case demonstrates the necessity of a passivity-oriented control structure,
that fundamentally mitigates the
negative-resistance property induced by the aforementioned intra-loop coupling.
If a controller can be designed to be dissipative
over a sufficiently wide frequency range, the inverter is more likely to
connect stably to a wide range of passive grid impedances.
Experimental verification of the observations in Subsections~\ref{Subsection:III-C} and \ref{Subsection:III-D} will be
provided in Section~\ref{Section:V}.

\section{Mitigation of Identified Negative-Resistance Mechanism via Parallel Virtual Resistance}
\label{Section:IV}

In this section, a method to enhance the passivity of VA-based GFM inverter is proposed,
informed by the identified negative-resistance mechanism of conventional VA-CC. 
It is shown that the negative-resistance property arising from the intra-loop coupling
can be effectively reduced by adding PVR to the VA, without degrading the
well-established CC or VFF,
and without requiring knowledge of the grid impedance.

\begin{figure}[!t]
    \centering
    \includegraphics[width=0.6\linewidth]{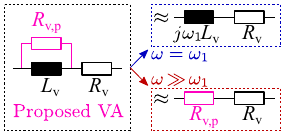}
    \caption{Diagram of proposed VA with PVR. 
    The PVR effect is negligible at the fundamental frequency but becomes dominant
    in the high-frequency range.}
    \label{fig:Diagram_PVR}
\end{figure}

\subsection{Mitigation of Control-Delay-Independent Non-Passivity from VA-CC via Parallel Virtual Resistance}

To avoid the negative-resistance property
induced by the \(s^2\)-term arising from the intra-loop coupling,
the VA-CC design should be reconsidered.
While the VA is typically designed as a series R-L circuit,
an actual physical inductor exhibits parallel dissipative components,
which introduce higher resistance in the high-frequency range.
This is observed as the skin effect, iron loss, and other frequency-dependent losses.

Likewise, if the VA exhibits a \emph{resistive} characteristic in the high-frequency range,
the negative-resistance property induced from the overall VA-CC---induced by the product of \(sL_\text{f}\) and \(sL_\text{v}\)---can be alleviated.
Inspired by this, the proposed method requires only adding a parallel dissipative component,
\(R_\text{v,p}\), to the original virtual inductance.
Its equivalent circuit diagram is shown in Fig.~\ref{fig:Diagram_PVR}.
Note that \(R_\text{v,p}\) is set large enough to have negligible effect at
the fundamental frequency,
thereby maintaining the inductive property in the fundamental and subsynchronous
frequency range.
Meanwhile, \(R_\text{v,p}\)
becomes significant in the harmonic range.
This suppresses the negative-resistance property induced by the \(s^2\)-term through two mechanisms:
\begin{enumerate}
  \item {In the high-frequency range, the admittance term \(1/R_\text{v,p}\) becomes dominant over \(1/(s L_\text{v})\).
Therefore, the \(180^\circ\) phase-shift response of \(s^2\), arising from the product of \(sL_\text{v}\) and \(sL_\text{f}\), can
be suppressed by reducing the phase-shift effect of \(sL_\text{v}\).}
  \item {The PVR \(R_\text{v,p}\) adds an additional resistive component;
specifically, \(\Re \left[ \frac{j \omega L_\text{v} R_\text{v,p}}{j \omega L_\text{v} + R_\text{v,p}} \right] \rightarrow R_\text{v,p}\)
as \(\omega \rightarrow \infty\). As discussed below, \(R_\text{v,p}\) is set significantly higher than \(\omega_1 L_\text{v}\)
(for example, \(R_\text{v,p} = \omega_\text{cc} L_\text{v}\)),
which provides a sufficient positive-real component in the harmonic range.}
\end{enumerate}

The new VA design is expressed as \(Y_\text{v}^\text{prop}(s)\):
\begin{equation}
  Y_\text{v}^\text{prop}(s)
  =\left( R_\text{v} + R_\text{v,p} ||  sL_\text{v} \right)^{-1}.
\end{equation}
The PVR should be designed to meet the necessary condition
that it does not introduce negative-resistance in the delay-absent condition.
Therefore, the delay-absent output impedance of the proposed VA-CC,
\(Z_\text{eq,0}^\text{prop} (s) \), is examined:
\begin{equation}
  \begin{aligned}
    Z_\text{eq,0}^\text{prop} (s) & = \frac{sL_\text{f}}{G_\text{cc} (s) Y_\text{v}^\text{prop}(s)}
    + \frac{1}{Y_\text{v}^\text{prop}(s)}
    \\ & \approx \frac{sL_\text{f}}{K_\text{p}Y_\text{v}^\text{prop}(s)} + \frac{1}{Y_\text{v}^\text{prop}(s)}
    \\ & = \frac{s }{\omega_\text{cc} }\left( \frac{s L_\text{v}R_\text{v,p}}{s L_\text{v} + R_\text{v,p}} + R_\text{v} \right) + \frac{s L_\text{v}R_\text{v,p}}{s L_\text{v} + R_\text{v,p}} + R_\text{v}.
  \end{aligned}
  \label{eq:Prop_Simple_Zeq}
\end{equation}
Substituting \(s= j\omega\), the real part is determined as
\begin{equation}
  \begin{aligned}
    & \Re \left[ \frac{j \omega }{\omega_\text{cc} }\left( \frac{j \omega L_\text{v}R_\text{v,p}}{j \omega L_\text{v} + R_\text{v,p}} + {R_\text{v}} \right)
    + \frac{j \omega L_\text{v}R_\text{v,p}}{j \omega L_\text{v} + R_\text{v,p}} + \underbrace{R_\text{v}}_{R_\text{v}>0} \right]
    \\ > &  \Re \left[ \frac{j \omega }{\omega_\text{cc} }\left( \frac{j \omega L_\text{v}R_\text{v,p}}{j \omega L_\text{v} + R_\text{v,p}} \right)
    + \frac{j \omega L_\text{v}R_\text{v,p}}{j \omega L_\text{v} + R_\text{v,p}}  \right]
    \\ = & \frac{\omega^2 L_\text{v} R_\text{v,p}}{\omega^2 L_\text{v}^2 + R_\text{v,p}^2} \underbrace{\left( L_\text{v} - \frac{R_\text{v,p}}{\omega_\text{cc}} \right)}_{\geq 0\text{: condition \eqref{eq:RvpCC_1}}}
  \end{aligned}
  \label{eq:Prop_Simple_Zeq_Real}
\end{equation}
Therefore, the condition to ensure positive real \(Z_\text{eq,0}^\text{prop} (s)\) is simply
\begin{equation}
  R_\text{v,p} \leq \omega_\text{cc} L_\text{v}.
  \label{eq:RvpCC_1}
\end{equation}
This holds for all \(\omega\),
satisfying the necessary condition under which the inner control loop does not introduce the
control-delay-independent non-passivity.
In the remainder of the manuscript, \(\omega_\text{cc} L_\text{v}\) is defined as \(R_\text{v,p}^\text{cc}\).

Note that setting \(R_\text{v,p}\) excessively small has a drawback,
as it affects the originally intended \(n_\text{XR}\) of the VA design.
Consequently, it is desirable to set \(R_\text{v,p}\) as high as possible, maintaining a sufficiently high
R/X ratio of the parallel R-L branch. For a parallel R-L branch,
a large R/X ratio corresponds to predominantly inductive behavior at the fundamental frequency,
in contrast to the series case where this role is played by the X/R ratio \(n_\text{XR}\).
The R/X ratio of the parallel branch is \( R_\text{v,p}/(\omega_1 L_\text{v})\).
Under the condition \(R_\text{v,p} = \omega_\text{cc} L_\text{v}\),
this parallel-branch R/X ratio becomes
\(\omega_\text{cc} / \omega_1\).
For a CC bandwidth of 1.5~kHz (25~p.u.), for example,
the R/X ratio is as high as 25,
which permits only a small adjustment
to the intended VA design at the fundamental frequency.
Therefore, it is recommended to set \(R_\text{v,p}\) as
\begin{equation}
  R_\text{v,p} = R_\text{v,p}^\text{cc} = \omega_\text{cc} L_\text{v}.
  \label{eq:RvpCC}
\end{equation}





\subsection{Evaluation of Proposed VA-CC with PVR under Practical Condition Considering Control Delay}


In this subsection, the simple design rule \eqref{eq:RvpCC} is examined
to determine whether it still provides a sufficiently wide positive-real
frequency range when the control delay is considered.
Note that the aim of the proposed method is to diminish the
negative-resistance property introduced by the intra-loop coupling within VA-CC.

It is worth mentioning that
the high-frequency negative-real region that emerges
due to the control delay itself is unavoidable for a practical inverter,
which has already been extensively investigated in past decade
\cite{Harnefors2014Passivity,Harnefors2016Passivity, Zou2018Analysis,Holmes2009Design, Agbemuko2021Passivity, Harnefors2015Passivity, Harnefors2017Nyquist}.
Nevertheless, for practical design,
the positive-real range does not need to extend to infinity.
Enforcing the positive-real condition up to an excessively high frequency
would unduly restrict the inverter's design flexibility.
Instead, a reasonable upper limit should be considered.
Two practical observations support this.
First, parasitic resistive components typically exhibit higher values
in the high-frequency range,
owing to the skin effect, iron loss, and other frequency-dependent loss mechanisms.
These naturally damp most of the potential instability \cite{Harnefors2017Nyquist}.
Second, the \(s\)-domain model used in the analysis does not accurately hold at frequencies approaching the
switching frequency, where switching effects, sampling, and discrete-time phenomena become significant.
Enforcing a positive-real condition in a regime where the underlying model itself is
unreliable therefore provides little practical benefit.
Therefore, a practical yet sufficient frequency range should be considered.
In this paper, the positive-real range is enforced up to 5~kHz,
which is one-quarter of the sampling frequency and half of the switching frequency.

\begin{figure}[!t]
    \centering
    \subfloat[]{
        \includegraphics[width=0.92\linewidth]{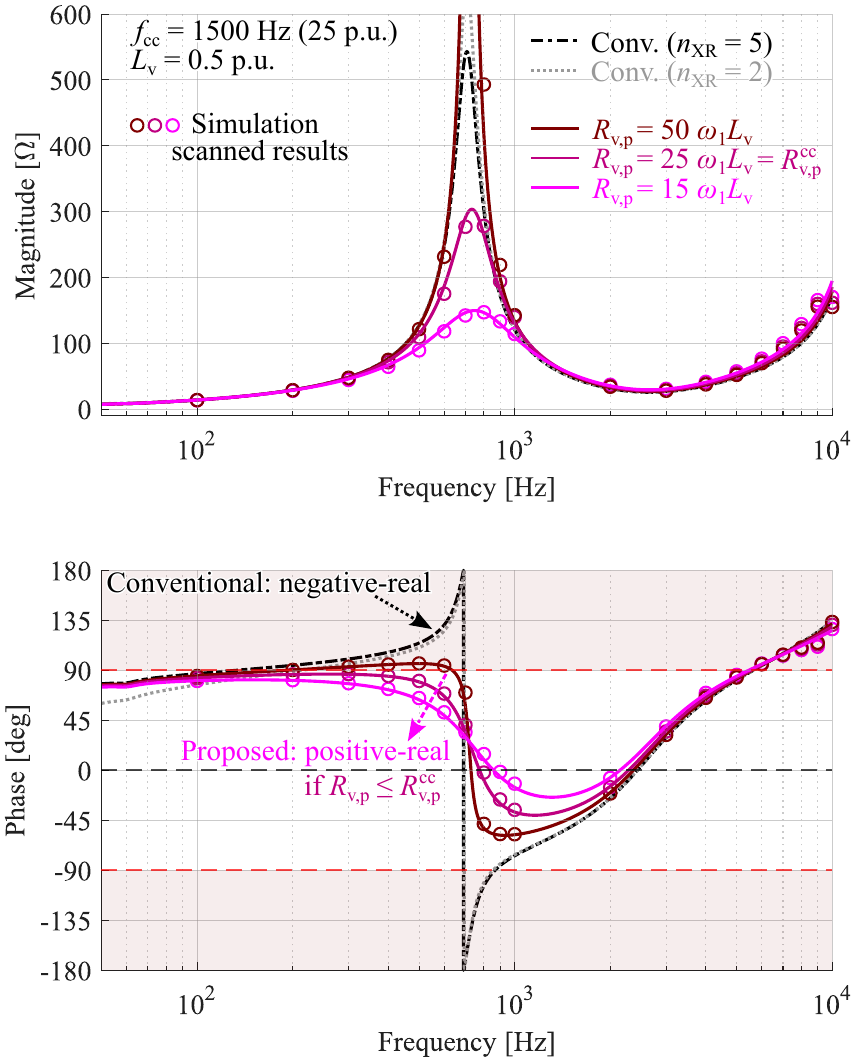}
    }
    \vspace{0pt}
    \subfloat[]{
        \includegraphics[width=0.92\linewidth]{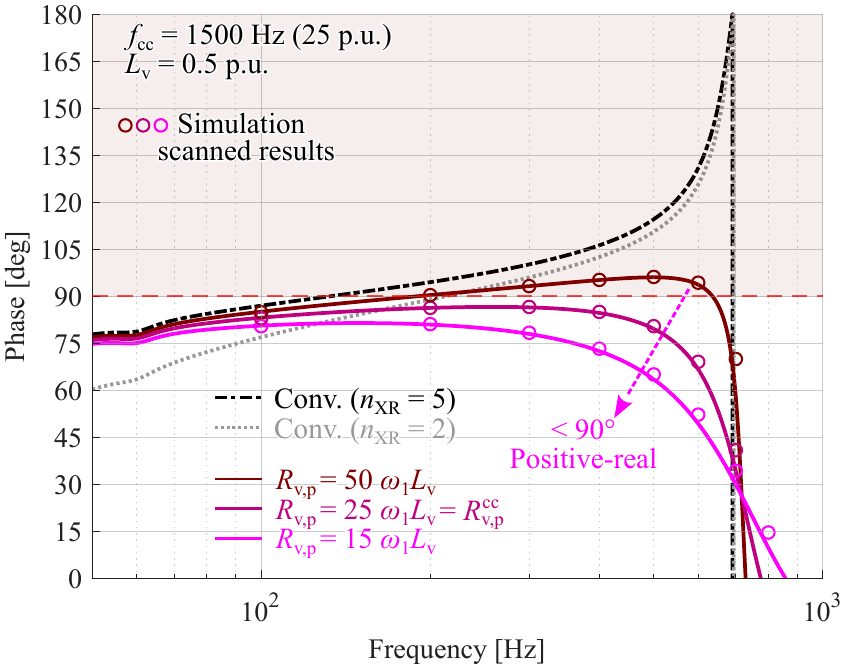}
    }
    \caption{Comparison of \(Z_\text{eq}(s)\) and \(Z_\text{eq}^\text{prop}(s)\) for different PVRs,
    together with the scanned results of the time-domain simulation.
    (a) From 50~Hz to 10~kHz. (b) Magnified view of the phase plot in (a), from 50~Hz to 1~kHz.}
    \label{fig:ValidationProposed}
\end{figure}

\begin{figure}[!t]
    \centering    
    \vspace{0pt}
    \subfloat[]{
        \includegraphics[width=0.47\linewidth]{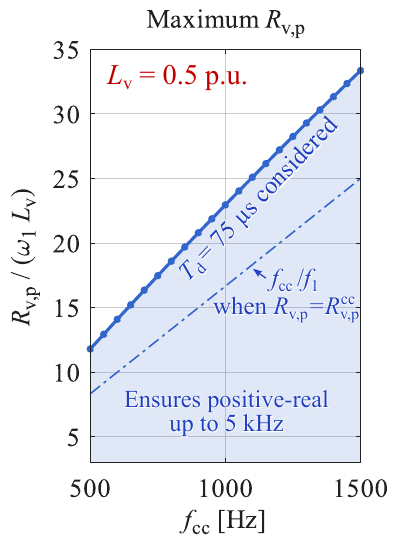}
    }
    \hfil
    \subfloat[]{
        \includegraphics[width=0.47\linewidth]{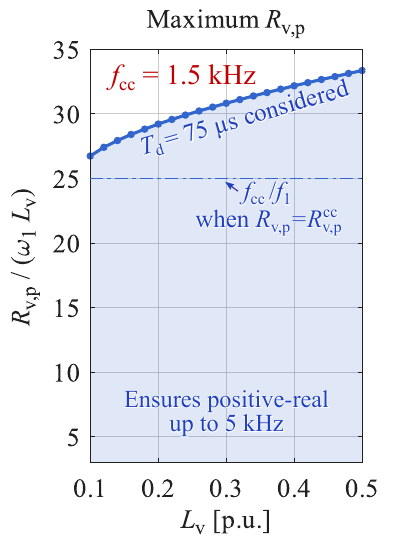}
    }
    \caption{Range of $R_\text{v,p}$  (shaded) that ensures a positive-real
    \(Z_\text{eq}^\text{prop}(s)\) up to 5~kHz, with \(T_\text{d} = 75~\mu\)s.
    (a) $L_\text{v}$ is fixed at 0.5~p.u. and $f_\text{cc}$ varies.
    (b) $f_\text{cc}$ is fixed at 1.5~kHz and $L_\text{v}$ varies.
    }
    \label{fig:ValidationProposed_RvMax}
\end{figure}

Considering the control delay, \(Z_\text{eq}^\text{prop}(s)\) is expressed as
\begin{equation}
  \begin{aligned}
  Z_\text{eq}^{\text{prop}}(s) & = - \frac{ \Delta \mathbf{v} }{ \Delta \mathbf{i} }
  = \frac{ G_\text{d}(s) G_\text{cc}(s) + sL_\text{f} }{1 - G_\text{d}(s) + G_\text{d}(s) G_\text{cc}(s) Y_\text{v}^\text{prop}(s)},
  \end{aligned}
  \label{eq:Full_Zeq_prop}
\end{equation}
similar to \eqref{eq:Full_Zeq},
but using \(Y_\text{v}^\text{prop}(s)\) instead of \(Y_\text{v}(s)\).
The impedance plots of \(Z_\text{eq}^\text{prop} (s)\) under different values of
\(R_\text{v,p}\) are shown in Fig.~\ref{fig:ValidationProposed}(a) and (b).
Fig.~\ref{fig:ValidationProposed}(b) provides a magnified view of the phase plot in Fig.~\ref{fig:ValidationProposed}(a) to better illustrate
that the proposed method effectively suppresses the
negative-resistance property introduced from conventional VA-CC.
As shown in Fig.~\ref{fig:ValidationProposed}(b).
with \(R_\text{v,p}\) equal to or less than \(R_\text{v,p}^\text{cc} = \omega_\text{cc}L_\text{v}\), 
the proposed method avoids the negative-resistance property below 1~kHz,
which is much favorable compared to conventional VA-CC.

For comparison, the conventional \(Z_\text{eq}(s)\) plots are also depicted.
Two cases with different series X/R ratios \(n_\text{XR}\) are shown to
illustrate that lowering \(n_\text{XR}\) does not generally help
mitigate the non-passivity.
In contrast, the proposed PVR only slightly affects the X/R ratio at the fundamental frequency
but effectively keeps \(Z_\text{eq}^\text{prop}(s)\) within the positive-real range across the
targeted harmonic frequencies.
Note that, due to the control delay, the negative-real region appears above approximately between 6 and 7~kHz,
which lies above the targeted 5~kHz range.

Whereas Fig.~\ref{fig:ValidationProposed} verifies the proposed method at a single design point,
Fig.~\ref{fig:ValidationProposed_RvMax}(a) and (b) sweep \(f_\text{cc}\) and \(L_\text{v}\) to
confirm that the condition in \eqref{eq:RvpCC} holds broadly across the design space.
The values of \(R_\text{v,p}\) for which \(Z_\text{eq}^\text{prop}(s)\) remains positive-real up to 5~kHz are numerically
calculated according to \eqref{eq:Full_Zeq_prop} for different values of \(f_\text{cc}\) and \(L_\text{v}\),
with a series-connected \(R_\text{v} = 0.2 \omega_1 L_\text{v}\).
The maximum value of \(R_\text{v,p}\) that achieves the positive-real
property is depicted in Fig.~\ref{fig:ValidationProposed_RvMax}(a) and (b).
In Fig.~\ref{fig:ValidationProposed_RvMax}(a), \(L_\text{v}\) is fixed at 0.5~p.u. and \(f_\text{cc}\) varies from
500~Hz to 1.5~kHz.
In Fig.~\ref{fig:ValidationProposed_RvMax}(b), the CC bandwidth is fixed at 1.5~kHz and \(L_\text{v}\) varies from
0.1~p.u. to 0.5~p.u.
The shaded area represents the range of \(R_\text{v,p}\) over which \(Z_\text{eq}^\text{prop}(s)\) remains positive-real
up to 5~kHz. In both cases, it is confirmed that \(R_\text{v,p}^\text{cc}\), determined by \eqref{eq:RvpCC}, keeps
\(Z_\text{eq}^\text{prop}(s)\) positive-real within the investigated frequency range.
Thus, \eqref{eq:RvpCC} provides a practically valid approach for designing the PVR.

\section{Experimental Verification}
\label{Section:V}



Experimental verification is performed on the scenarios established in Section~\ref{Section:III} and \ref{Section:IV},
using the same default parameters listed in Table~\ref{tab:SectionIII}.
Each scenario demonstrates the instability conditions of conventional VA-CC,
along with the mitigation provided by the proposed PVR.
It is demonstrated that, since the proposed PVR ensures
positive-real property up to a sufficiently high-frequency range,
it promotes stability under a practical passive grid condition.




\begin{figure*}[!t]
    \centering
    \subfloat[]{
        \includegraphics[width=0.98\linewidth]{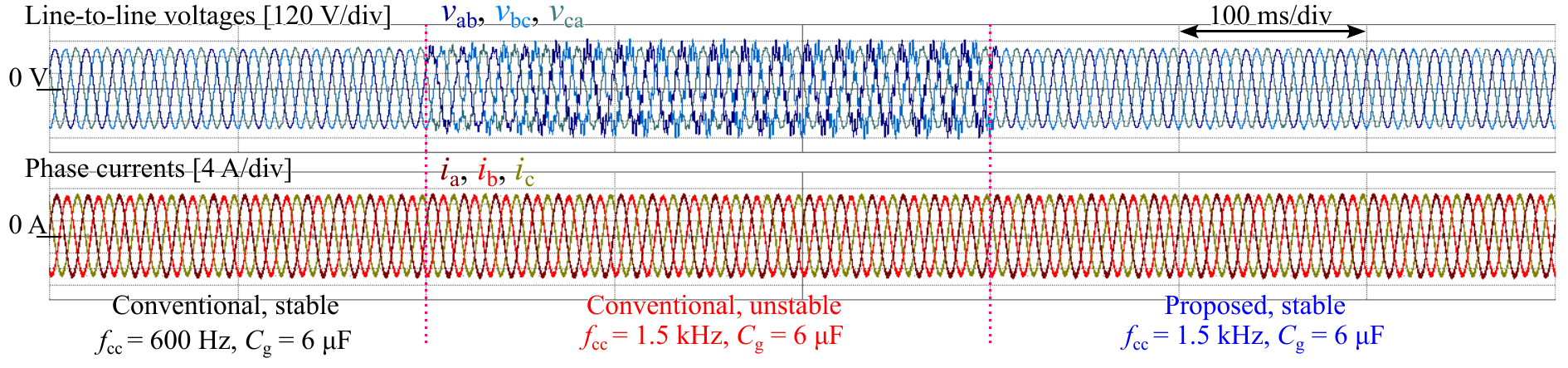}
    }
    \vspace{-15pt}
    \subfloat[]{
        \includegraphics[width=0.45\linewidth]{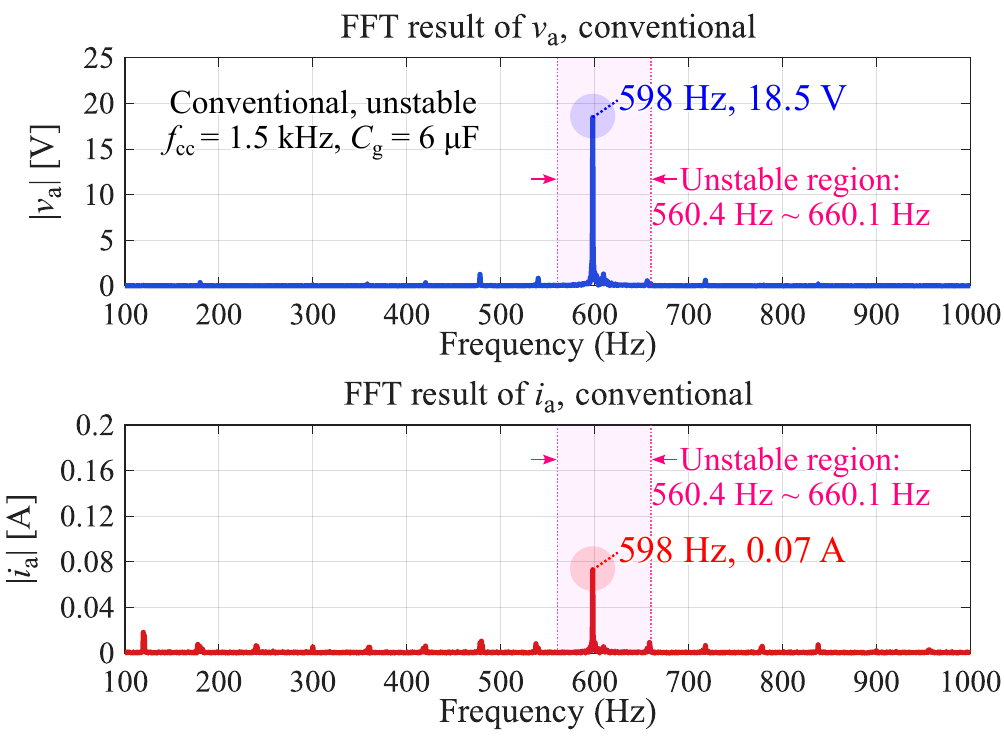}
    }
    \hfil{}
    \subfloat[]{
        \includegraphics[width=0.45\linewidth]{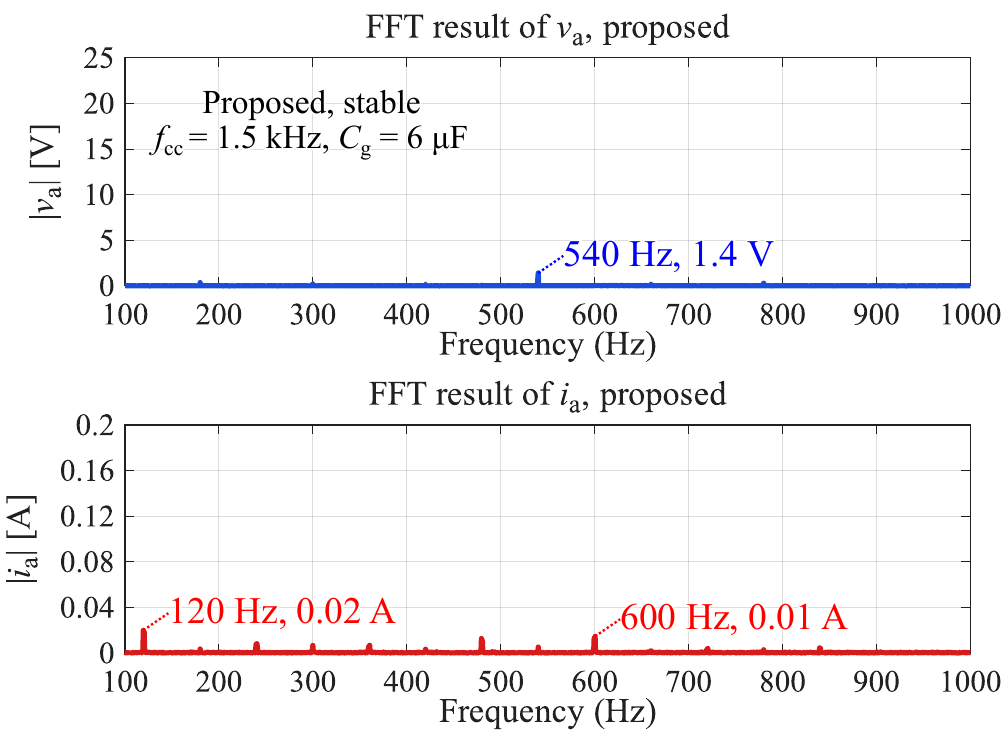}
    }
    \caption{Experimental verification of Subsection~\ref{Subsection:III-C}. (a) Experimental results, from the conventional method ($ f_\text{cc} = 600 $~Hz to $ f_\text{cc} = 1.5 $~kHz) to the proposed method ($ f_\text{cc} = 1.5 $~kHz).
    (b) FFT result of the conventional method, (c) FFT result of the proposed method.}
    \label{fig:Experiment1_FFT}
\end{figure*}

\subsection{Effect of CC Bandwidth}

The effects of different CC bandwidth \(f_\text{cc}\) are compared in Fig.~\ref{fig:Experiment1_FFT}.
The CC bandwidth of conventional VA-CC is increased from 600~Hz to 1.5~kHz under the
same conditions stated in Subsection~\ref{Subsection:III-C}.
In the first stage of Fig.~\ref{fig:Experiment1_FFT}(a), the controller remains stable with \(f_\text{cc} = 600\)~Hz,
but becomes unstable in the second stage with \(f_\text{cc} = 1.5\)~kHz,
as predicted in Subsection~\ref{Subsection:III-C}.
Finally, the proposed PVR of
\(R_\text{v,p}^\text{cc} = 2 \pi \times 1500 \times L_\text{v}\)
is added
without changing the CC bandwidth of 1.5~kHz.
As expected, the harmonic instability is effectively suppressed.
This highlights an additional benefit of the proposed approach.
In the proposed method,
the CC bandwidth can be selected according to the sampling frequency,
rather than being constrained by the structural
non-passivity problem from conventional VA-CC.
The proposed method thus allows the controller to
fully exploit a high sampling frequency,
towards higher CC bandwidth.

For further verification of the theoretical analysis,
FFT result of the conventional and proposed methods are compared
in Fig.~\ref{fig:Experiment1_FFT}(b) and (c).
According to Fig.~\ref{fig:Validation4_CC},
unstable frequency region arises between 560.4~Hz and 660.1~Hz.
The FFT result in Fig.~\ref{fig:Experiment1_FFT}(b) matches this analysis,
as the main
oscillation frequency is 598~Hz. 
Meanwhile, as shown in Fig.~\ref{fig:Experiment1_FFT}(c),
the proposed method remains stable,
and no severe oscillation occurs.

\begin{figure*}[!t]
    \centering
    \subfloat[]{
        \includegraphics[width=0.98\linewidth]{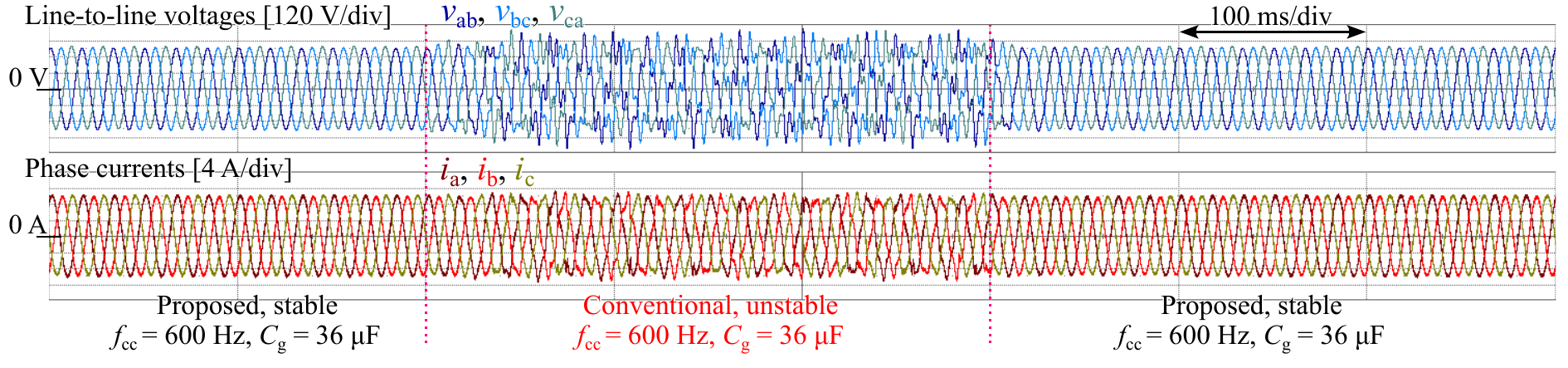}
    }
    \vspace{-15pt}
    \subfloat[]{
        \includegraphics[width=0.45\linewidth]{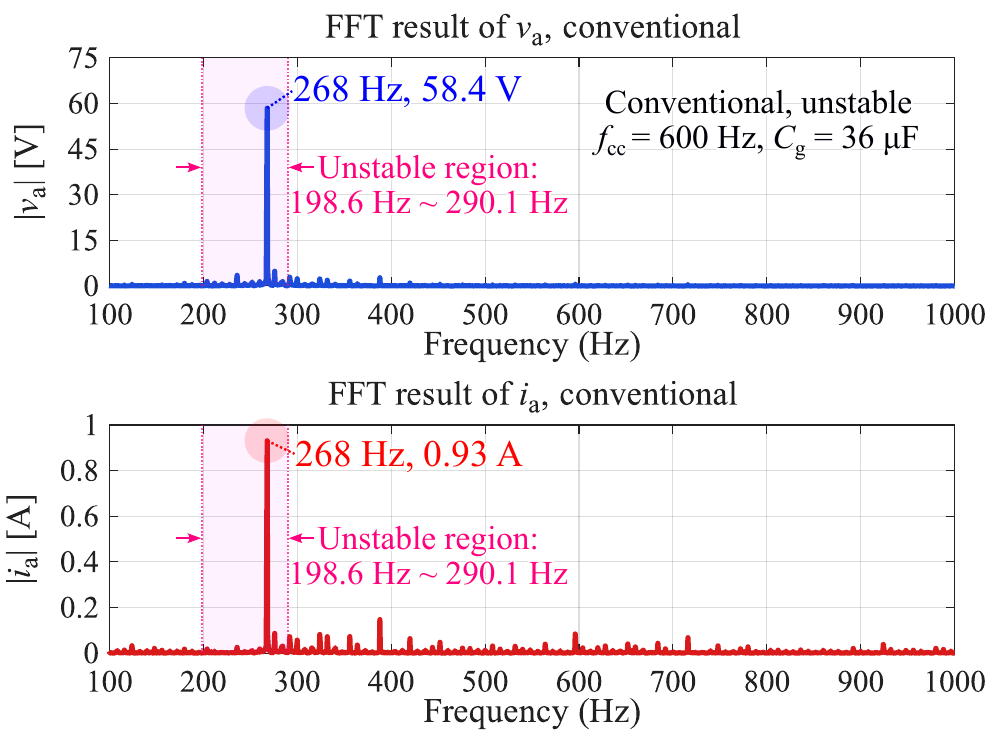}
    }
    \hfil{}
    \subfloat[]{
        \includegraphics[width=0.45\linewidth]{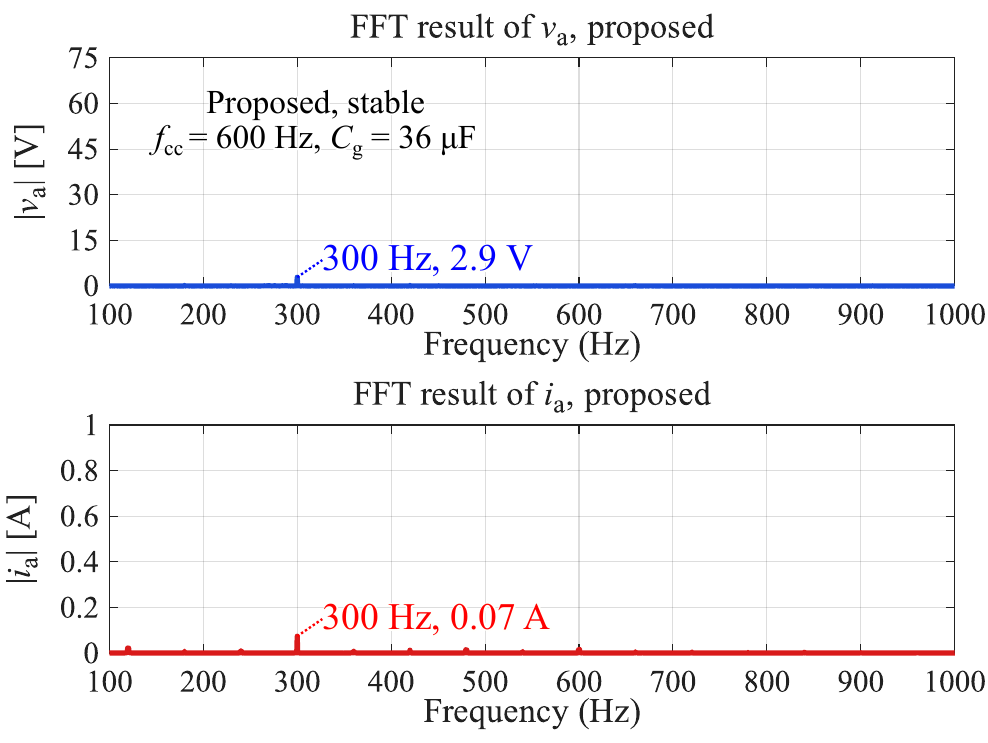}
    }
    \caption{Experimental verification of Subsection~\ref{Subsection:III-D}, with $ C_\text{g} = 36~\mu$F.
    (a) Experimental results, from the proposed method to conventional method, then back to proposed method ($ f_\text{cc} = 600 $~Hz).
    (b) FFT result of the conventional method, (b) FFT result of the proposed method.}
    \label{fig:Experiment2_FFT}
\end{figure*}

\begin{figure*}[!t]
    \vspace{0pt}
    \centering
    \includegraphics[width=0.95\linewidth]{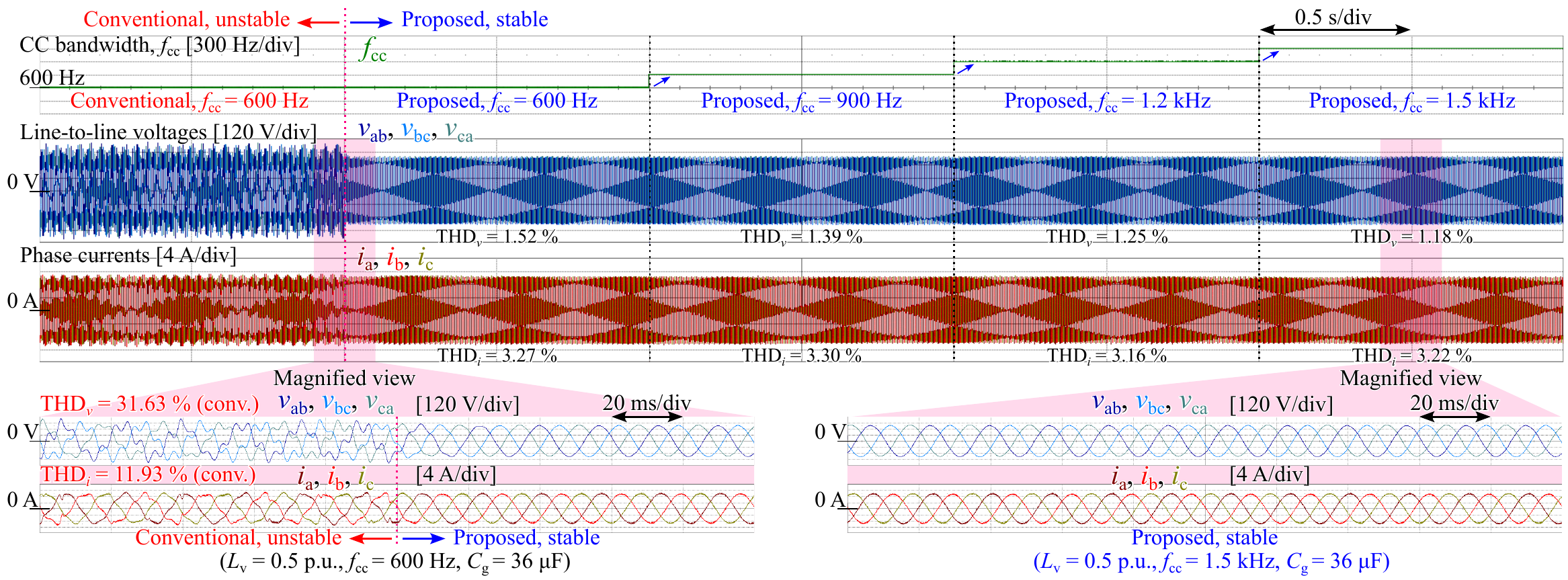}
    \caption{Experimental verification of the proposed method under various CC bandwidths,
    under the grid condition of SCR\(=4\) and \(C_\text{g} = 36~\mu\)F.
    Initially, conventional method is applied, then switched to the proposed method.
    \(f_\text{cc}\) is changed in a stepwise manner: 600~Hz, 900~Hz, 1.2~kHz, and 1.5~kHz.}
    \label{fig:proposed_fccstep}
\end{figure*}

\subsection{Effect of the Grid Resonant Frequency}

The undesirable grid condition from Subsection~\ref{Subsection:III-D} is tested, in which
the grid-side LC resonant frequency
\(f_\text{LC,g}\) lies below \(f_D\) of conventional VA-CC.
The result is provided in Fig.~\ref{fig:Experiment2_FFT}.
Since the conventional method
cannot ensure stability under the severe grid condition
SCR\(=\)4 and \(C_\text{g}=36~\mu\)F,
the experiment begins with the
proposed PVR already active.
In the first stage of Fig.~\ref{fig:Experiment2_FFT}(a), a PVR of
\(R_\text{v,p}^\text{cc} = \omega_\text{cc} L_\text{v}\)
with \(f_\text{cc} = 600\)~Hz
is applied,
and the inverter operates stably despite the unfavorable grid condition.
In the second stage, the PVR is deactivated,
restoring the conventional VA-CC.
The harmonic instability emerges, with the oscillation appearing near \(f_\text{LC,g}\),
as predicted by the analysis in Subsection~\ref{Subsection:III-D}.
In the last stage, the PVR is re-engaged,
and the system promptly returns to stable operation,
which confirms the effectiveness of the proposed PVR.

\begin{figure*}[!t]
    \centering
    \includegraphics[width=0.95\linewidth]{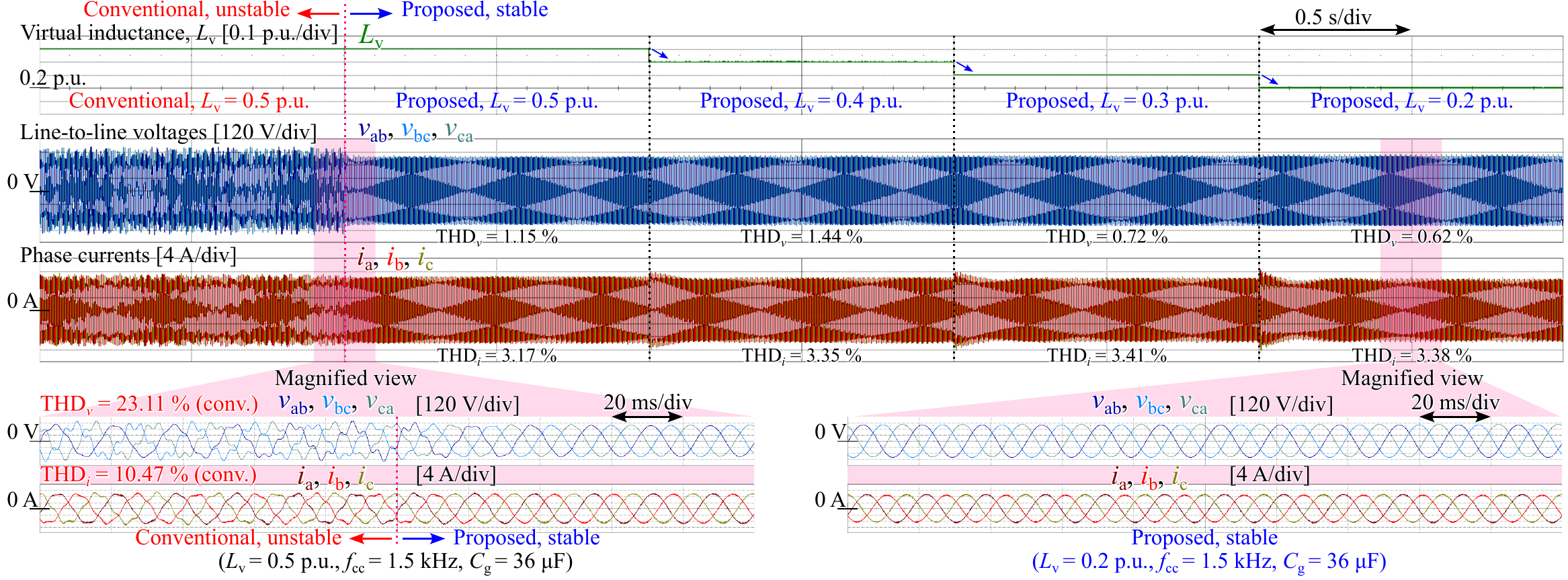}
    \caption{Experimental verification of the proposed method under various \(L_\text{v}\),
    under the grid condition of SCR\(=4\) and \(C_\text{g} = 36~\mu\)F.
    Initially, conventional method is applied, then switched to the proposed method.
    \(L_\text{v}\) is changed in a stepwise manner: 0.5~p.u., 0.4~p.u., 0.3~p.u., and 0.2~p.u. In all cases, \(n_\text{XR}=5\).}
    \label{fig:proposed_Lvstep}
\end{figure*}

The FFT results for the case \(C_\text{g} = 36~\mu\)F are also provided in
Fig.~\ref{fig:Experiment2_FFT}(b) and (c).
According to Fig.~\ref{fig:Validation4_Grid},
an unstable frequency region arises between 198.6~Hz and 290.1~Hz.
The FFT result in Fig.~\ref{fig:Experiment2_FFT}(b) shows the main oscillation occurring at 268~Hz,
which is well within the expected range in Fig.~\ref{fig:Validation4_Grid}.
Meanwhile, the proposed method shows no significant
oscillation as shown in Fig.~\ref{fig:Experiment2_FFT}(c).



\subsection{Evaluation of the Proposed Method over Various Control Parameters}

Finally, the proposed method is further evaluated over various control parameters.
In all cases, the undesirable grid condition with the low \(f_\text{LC,g}\) is considered,
i.e., SCR\(=\)4 and \(C_\text{g}=36~\mu\)F.
First, in Fig.~\ref{fig:proposed_fccstep}, the conventional method is initially applied and then switched to
the proposed method.
Under the proposed method,
\(f_\text{cc}\) is increased from 600~Hz to 1.5~kHz in a stepwise manner.
In each \(f_\text{cc}\) cases, the PVR of \(R_\text{v,p}^\text{cc} = \omega_\text{cc} L_\text{v}\) is applied,
which ensures positive-real property up to a sufficiently high frequency range, as shown in Fig.~\ref{fig:ValidationProposed_RvMax}(a).
Unlike the conventional method, the proposed method remains stable across a wide range of CC bandwidths.
Therefore, it can fully exploit the high sampling frequency of 20~kHz
to achieve a higher CC bandwidth
despite the unfavorable grid condition,
without inducing harmonic instability.

Since \(R_\text{v,p}^\text{cc} = \omega_\text{cc} L_\text{v}\) also depends on
\(L_\text{v}\), further evaluation on different \(L_\text{v}\) is required.
Fig.~\ref{fig:proposed_Lvstep} validates the case with various \(L_\text{v}\).
Note that, although sufficient \(L_\text{v}\) may be favorable for the stable operation of GFM inverter,
Fig.~\ref{fig:proposed_Lvstep} intentionally considers the cases with \(L_\text{v}\) lower than 0.5~p.u.
for evaluation purpose.
The CC bandwidth is fixed at the relatively high \(f_\text{cc} =1.5\)~kHz in all cases.
For each \(L_\text{v}\) cases, PVR of
\(R_\text{v,p}^\text{cc} = \omega_\text{cc} L_\text{v}\) is applied.
As \(L_\text{v}\) is decreased in a stepwise manner, the current transients appear at each
transition. Nevertheless, the proposed method reaches stable steady-state condition,
as the equivalent output impedance
\(Z_\text{eq}^\text{prop}(s)\)
preserves positive-real property shown in Fig.~\ref{fig:ValidationProposed_RvMax}(b).

\section{Conclusion}
\label{Section:VI}

This paper has identified a negative-resistance property in output impedance that originates
within VA-CC itself, regardless of the control delay.
Unlike previous works that regard the control delay as the main cause
of non-passivity in the harmonic range,
this work traces the mechanism to an
\(s^2\)-term in the equivalent output impedance,
introduced by the intra-loop coupling among the constituents of VA-CC:
VA, CC, and VFF.
The \(s^2\)-term induces a negative-resistance property in a control-delay-independent manner,
and renders the VA-CC-based GFM inverter non-passive in the harmonic range.
Consequently, it readily triggers harmonic instability when the grid contains a shunt capacitance.
To address this structural shortcoming at its root,
a passivity-oriented design is proposed based on the identified mechanism.
A PVR is introduced into the VA,
with a simple and explicit design rule, \(R_\text{v,p}^\text{cc} = \omega_\text{cc} L_\text{v}\).
The modified VA-CC thereby mitigates the negative-resistance property
arising from the intra-loop coupling,
while fully retaining the well-established CC and VFF,
and without requiring grid impedance information.
The analysis and the proposed mitigation method are validated through experiments
in the scenarios that expose the failure of conventional VA-CC.

\section{Appendix}
\label{Section:Appendix}

This appendix provides the detailed reasoning for the two observations
stated in Subsection~\ref{Subsection:III-B}:
(a) \(\omega_D < \omega_N\),
and (b) the denominator \(D(s)\) is typically underdamped
under practical parameter design.

Comparing \eqref{eq:omega_N} and \eqref{eq:omega_D},
\begin{equation}
  \begin{aligned}
  \omega_N/\omega_D = \sqrt{2L_\text{v} / L_\text{f}}.
  \end{aligned}
\end{equation}
The VA is designed to enhance stability of the GFM inverter by adding inductance,
and therefore \(L_\text{v} \geq L_\text{f}\) is employed.
Thus, \(2L_\text{v} > L_\text{f}\) is naturally satisfied, and
\(\omega_D < \omega_N\) can be assumed.

The damping coefficient \(\zeta_{D}\) of the denominator \(D(s)\) is
\begin{equation}
  \begin{aligned}
  \zeta_{D} = & \,\  \left( \frac{ R_\text{v}}{2} - \frac{ K_\text{p}}{4} \right) \sqrt{\frac{T_\text{d}}{L_\text{v} K_\text{p}}}.
  \end{aligned}
\end{equation}
Note that
\( \left| \zeta_{D} \right| = \left|\left( \frac{R_\text{v}}{2} - \frac{K_\text{p}}{4} \right) \sqrt{\frac{T_\text{d} }{L_\text{v} K_\text{p}}} \right| \), 
and therefore
\begin{equation}
\left| \zeta_{D} \right| <
\max \left(\frac{R_\text{v}}{2} \sqrt{\frac{T_\text{d} }{ L_\text{v} K_\text{p}}} , \frac{K_\text{p}}{4} \sqrt{\frac{T_\text{d} }{ L_\text{v} K_\text{p}}} \right).  
\end{equation}
Consider an upper bound for \(|\zeta_D|\) in the most conservative parameter conditions.
For the first term \( \frac{R_\text{v}}{2} \sqrt{\frac{T_\text{d} }{ L_\text{v} K_\text{p}}} \),
this is equal to
\begin{equation}
\frac{ \omega_1 }{2 n_\text{XR}} \sqrt{\frac{ T_\text{d} L_\text{v} }{ \omega_\text{cc} L_\text{f}}}.
\end{equation}
\begin{enumerate}
  \item \(L_\text{v}\) should be maintained high for sufficient \(\zeta_D\),
  but it must be smaller than 1~p.u. to preserve the margin for transient stability.
  The extreme case \(L_\text{v} \leq 1\)~p.u. is therefore considered.
  \item The minimum value of \(L_\text{f}\) depends on the topology, control delay, and other factors,
  but it is typically larger than 0.05~p.u. Thus, \(L_\text{f} \geq 0.05\)~p.u. is considered.
  \item The lowest possible \(n_\text{XR}\) is considered, while keeping the impedance inductive:
  \(n_\text{XR} \geq 2\).
  \item Consider \(\omega_\text{cc}\) as low as possible while remaining sufficiently faster than the outer loops to regulate the fundamental component:
  \(\omega_\text{cc} \geq 3 \omega_1\).
  \item Most critically, the control delay should be sufficiently small to protect the inverter during faults.
  Here, a sampling frequency of 10~kHz is considered, giving \(T_\text{d} = 1.5/f_\text{s} = 150~\mu\text{s}\).
\end{enumerate}
Even under the most conservative parameter conditions listed above,
this bound on \(|\zeta_D|\) evaluates to 0.1535 in a 60~Hz system.

For the second term \(\frac{T_\text{d} K_\text{p}}{4} \sqrt{\frac{1}{T_\text{d} L_\text{v} K_\text{p}}} = \frac{1}{4}\sqrt{ \frac{T_\text{d} \omega_\text{cc} L_\text{f}}{L_\text{v}}}\),
the conservative bounds are as follows:
\begin{enumerate}
  \item The CC bandwidth is limited by \(\omega_\text{cc} \leq 2\pi f_\text{s} /10\) when \(T_\text{d} = 1.5/f_\text{s}\);
  thus, \(\omega_\text{cc} \leq 1 / T_\text{d} \) is considered.
  \item \(L_\text{v}\) is set larger than \(L_\text{f}\) to add virtual inductance
  to the line impedance; thus, \(L_\text{f} \leq L_\text{v}\) is taken.
\end{enumerate}
Even under the most conservative case listed above,
\(\frac{1}{4}\sqrt{ \frac{T_\text{d} \omega_\text{cc} L_\text{f}}{L_\text{v}}}\)
evaluates to a maximum of 0.25.


\end{document}